\documentclass[twocolumn]{aastex631}
\usepackage{booktabs}
\usepackage{multirow}
\usepackage{graphicx}
\usepackage{amsmath} 

\providecommand{\teff}{\ensuremath{T_{\rm eff}}} 
\providecommand{\msun}{\ensuremath{\,{\rm M_{\odot}}}} 
\providecommand{\rsun}{\ensuremath{\,{\rm R_{\odot}}}} 
\providecommand{\vsini}{\ensuremath{\,{\rm v\sin{i_\star}}}} 

\shorttitle{TOI-880 is an Aligned, Coplanar, Multi-planet System}
\shortauthors{Zhang et al.}

\begin{document}

\title{TOI-880 is an Aligned, Coplanar, Multi-planet System}

\author[0009-0004-0455-2424]{Elina Y. Zhang}
\affiliation{Institute for Astronomy, University of Hawaii, 2680 Woodlawn Drive, Honolulu, HI 96822 USA}

\author[0000-0003-3860-6297]{Huan-Yu Teng}
\affiliation{Korea Astronomy and Space Science Institute, 776 Daedeok-daero, Yuseong-gu, Daejeon 34055, Republic of Korea}
\affiliation{National Astronomical Observatories, Chinese Academy of Sciences, Beijing 100101, China}
\affiliation{Institute for Astronomy, University of Hawaii, 2680 Woodlawn Drive, Honolulu, HI 96822 USA}

\author[0000-0002-8958-0683]{Fei Dai}
\affiliation{Institute for Astronomy, University of Hawaii, 2680 Woodlawn Drive, Honolulu, HI 96822 USA}

\author[0000-0001-8638-0320]{Andrew W. Howard} 
\affiliation{Department of Astronomy, California Institute of Technology, Pasadena, CA 91125, USA}

\author[0000-0003-1312-9391]{Samuel P. Halverson} 
\affiliation{Jet Propulsion Laboratory, California Institute of Technology, 4800 Oak Grove Drive, Pasadena, CA 91109, USA}

\author[0000-0002-0531-1073]{Howard Isaacson} 
\affiliation{{Department of Astronomy,  University of California Berkeley, Berkeley CA 94720, USA}}

\author[0000-0003-3856-3143]{Ryan A. Rubenzahl}  
\affiliation{Center for Computational Astrophysics, Flatiron Institute, 162 Fifth Avenue, New York, NY 10010, USA}

\author[0000-0002-0376-6365]{Xian-Yu Wang}
\affiliation{Department of Astronomy, Indiana University, Bloomington, IN 47405, USA}

\author[0000-0002-7846-6981]{Songhu Wang}
\affiliation{Department of Astronomy, Indiana University, Bloomington, IN 47405, USA} 

\author[0000-0003-3504-5316]{Benjamin J. Fulton} 
\affiliation{NASA Exoplanet Science Institute/Caltech-IPAC, MC 314-6, 1200 E California Blvd, Pasadena, CA 91125, USA}

\author[0000-0002-5254-2499]{Louise D. Nielsen} 
\affiliation{University Observatory, Faculty of Physics, Ludwig-Maximilians-Universit{\"a}t M{\"u}nchen, Scheinerstr. 1, 81679 Munich, Germany}

\author[0000-0001-8342-7736]{Jack Lubin} 
\affiliation{Department of Physics \& Astronomy, University of California Los Angeles, Los Angeles, CA 90095, USA}

\author[0000-0002-8965-3969]{Steven Giacalone} 
\affiliation{Department of Astronomy, California Institute of Technology, Pasadena, CA 91125, USA}

\author[0000-0002-9305-5101]{Luke B. Handley} 
\affiliation{Department of Astronomy, California Institute of Technology, Pasadena, CA 91125, USA}

\author[0000-0003-0967-2893]{Erik A. Petigura} 
\affiliation{Department of Physics \& Astronomy, University of California Los Angeles, Los Angeles, CA 90095, USA}

\author[0000-0002-1845-2617]{Emma V. Turtelboom}
\affiliation{Department of Astronomy, 501 Campbell Hall, University of California, Berkeley, CA 94720, USA}

\author[0000-0001-7047-8681]{Alex S. Polanski}
\affiliation{Department of Physics and Astronomy, University of Kansas, Lawrence, KS 66045, USA}

\author[0009-0004-4454-6053]{Steve R. Gibson} 
\affil{Caltech Optical Observatories, Pasadena, CA, 91125, USA}

\author{Kodi Rider} 
\affil{Space Sciences Laboratory, University of California Berkeley, Berkeley, CA 94720, USA}

\author[0000-0001-8127-5775]{Arpita Roy} 
\affiliation{Astrophysics \& Space Institute, Schmidt Sciences, New York, NY 10011, USA} 

\author[0000-0002-6525-7013]{Ashley Baker} 
\affil{Caltech Optical Observatories, Pasadena, CA, 91125, USA}

\author[0009-0002-2419-8819]{Jerry Edelstein} 
\affil{Space Sciences Laboratory, University of California Berkeley, Berkeley, CA 94720, USA}

\author{Christopher L. Smith} 
\affil{Space Sciences Laboratory, University of California Berkeley, Berkeley, CA 94720, USA}

\author[0000-0002-6092-8295]{Josh Walawender} 
\affiliation{W. M. Keck Observatory, 65-1120 Mamalahoa Hwy, Waimea, HI 96743}

\author[0000-0002-4265-047X]{Joshua N. Winn} 
\affiliation{Department of Astrophysical Sciences, Princeton University, 4 Ivy Lane, Princeton, NJ 08544, USA} 

\begin{abstract}
\noindent Although many cases of stellar spin-orbit misalignment are known, it is usually unclear whether a single planet's orbit was tilted or if the entire protoplanetary disk was misaligned. Measuring stellar obliquities in multi-transiting planetary systems helps to distinguish these possibilities. Here, we present a measurement of the sky-projected spin-orbit angle for TOI-880 c (TOI-880.01), a member of a system of three transiting planets, using the Keck Planet Finder (KPF). We found that the host star is a K-type star ($\teff=5050 \pm 100$ K). Planet b (TOI-880.02) has a radius of $2.19\pm0.11\mathrm{R_{\oplus}}$ and an orbital period of $2.6$ days; planet c (TOI-880.01) is a Neptune-sized planet with $4.95\pm0.20\mathrm{R_{\oplus}}$ on a $6.4$-day orbit; and planet d (TOI-880.03) has a radius of $3.40_{-0.21}^{+0.22}\mathrm{R_{\oplus}}$ and a period of $14.3$ days. By modeling the Rossiter-McLaughlin (RM) effect, we found the sky-projected obliquity to be $|\lambda_c| = 7.4_{-7.2}^{+6.8}$$^{\circ}$, consistent with a prograde, well-aligned orbit. The lack of detectable rotational modulation of the flux of the host star and a low $\vsini$ (1.6~km/s) imply slow rotation and correspondingly slow nodal precession of the planetary orbits and the expectation that the system will remain in this coplanar configuration. TOI-880 joins a growing sample of well-aligned, coplanar, multi-transiting systems. Additionally, TOI-880 c is a promising target for JWST follow-up, with a transmission spectroscopy metric (TSM) of $\sim 170$. We could not detect clear signs of atmospheric erosion in the H$\alpha$ line from TOI-880 c, as photoevaporation might have diminished for this mature planet.
\end{abstract}


\section{Introduction} 
\label{sec:intro}
The stellar obliquity of a planet is a clue about its dynamical history. The stellar obliquity $\psi$ is the angle between a planet's orbital axis ($\hat{n}_{orb}$) and the host star's spin axis ($\hat{n}_{\star}$). The sky-projected obliquity, $\lambda$, is the angle between the sky projections of $\hat{n}_{orb}$ and $\hat{n}_{\star}$. The Rossiter-McLaughlin \citep[RM;][]{Rossiter1924, McLaughlin1924} effect is a spectroscopic anomaly observed during planetary transits that allows
for a measurement of the sky-projected obliquity. When a planet transits its host star, it blocks part of the rotating stellar photosphere. When the planet is in front of the approaching (blue-shifted) half of the star, the disk-integrated starlight appears slightly red-shifted, and vice versa. The result is an “anomalous Doppler shift” that varies over time throughout the transit. The amplitude of the anomalous Doppler shift, expressed as a radial velocity, is
\vspace{-0.5em}
\begin{equation}
    \Delta RV_{max} \sim 15~{\rm m\ s}^{-1} \left(\frac{R_p/R_{Jup}}{R_{\star}/R_{\odot}} \right)^2 \left( \frac{v\sin{i_\star}}{2~{\rm km\ s}^{-1}} \right)
\vspace{-0.5em}
\end{equation} 
where $\rm v\sin{i_\star}$ is the line-of-sight projection of the stellar rotation velocity. If the inclination of the stellar spin axis $i_{\star}$ is also known, e.g. from astereoseismology \citep{Kepler-56,Ong2024} or stellar rotational broadening \citep{Albrecht2021}, one can constrain the true stellar obliquity $\psi$.

The solar system is aligned with the Sun, as the planetary orbits lie near the invariable plane, inclined by $\sim6^\circ$ to the solar equator \citep{Souami2012, Beck2005}. However, this is not always the case for exoplanetary systems. Diverse configurations have been found, including prograde, polar, and possibly retrograde orbits \citep{Albrecht2022}. Previous studies have revealed several trends about stellar obliquity. \citet{Winn2010} proposed a difference in the obliquity distributions of hot Jupiters between hot host stars and cool stars, with cool stars tending to have low obliquities, and hot stars having a broad range of obliquities. However, for warm Jupiters, a recent study by \citet{Wang2024} stated that single-star systems with warm Jupiters tend to be aligned, even when the host star is hot. \citet{Hebrard2011} and more recently \citet{Rusznak2024} suggested a relationship between obliquity with stellar and planetary masses, stating that stars with especially massive planets have lower obliquities. Also, a relationship between obliquity and eccentricity has been proposed by
\citet{Rice2022}, declaring that cool stars with eccentric planets tend to have higher obliquities than those with planets on nearly circular orbits.

NASA’s {\it Kepler} \citep{Borucki} and {\it TESS} \citep{Ricker} missions have discovered many systems with multiple transiting planets. The misaligned orbit of a single transiting hot Jupiter is often attributed to high-eccentricity migration or some other dynamical process that tilted the orbit of a single planet \citep{Winn2010, Dawson}. A priori, a multi-transiting system likely has low mutual inclinations between the planets; otherwise, one observer is unlikely to observe multiple transiting planets. Therefore, an obliquity measurement of a multi-transiting system directly tests theories that tilt the whole planetary system, such as protoplanetary disk tilting \citep{Lai2011, Lai2014}, or mechanisms that cause the host star's spin axis to tumble randomly \citep{Rogers2012, Rogers2013}, against those that excite mutual inclinations between planets, such as secular chaos \citep{WuLithwick2011}. Up to now, there are about 25 multi-transiting systems that have obliquity measurements, and most of them have low obliquities \citep{Albrecht2022}. This trend hints that multi-transiting systems are dynamically cold in general. However, there are three reports of misaligned systems: HD 3167 \citep{HD3167}, Kepler-56 \citep{Kepler-56}, and K2-290 \citep{K2-290}. In addition, AU Mic c shows a potential misalignment ($\lambda=68^{+32}_{-49}$$^{\circ}$) but remains consistent with 0 degrees within 1.5$\sigma$ \citep{AU_mic_c}. Recently, \citet{Teng2025} cast doubt on the previously reported polar orbit of HD~3167~c \citep{HD3167}, advocating for additional obliquity measurements to better constrain its orbital architecture. \citet{Ong2024} re-examined the Kepler-56 system and proposed that the star's core and envelope rotate around different axes: the core is misaligned with the planets, whereas the envelope is aligned with them. Following these developments, K2-290 \citep{K2-290} remains a multi-transiting system with a misaligned orbit; however, it is important to note that it is in a multi-star system.

In this paper, we present the stellar obliquity measurement of another multi-transiting system, TOI-880 (TIC 34077285), which hosts three transiting planets. We observed the Rossiter–McLaughlin (RM) effect of TOI-880 c (TOI-880.01) with the Keck Planet Finder (KPF), and analyzed the data jointly with {\it TESS} light curves. Section \ref{sec:observation} describes the {\it TESS} photometry and KPF spectroscopic observations. Section \ref{sec:stellar_para} explains the spectroscopic analysis of the host star’s parameters. Section \ref{sec:analysis} presents our joint analysis of the {\it TESS} light curves and the RM data, along with the resulting stellar obliquity. Section \ref{sec:mul-tran} and \ref{sec:precession} describe the dynamical modeling of the system. Section \ref{sec:atm} investigates the potential atmospheric loss. Section \ref{sec:discussion} discusses our results, and Section \ref{sec:summary} summarizes the key findings.

\section{Observations} \label{sec:observation}

\subsection{{\it TESS} Light Curves} 
TOI-880 was observed by the {\it TESS} mission in Sector 6 at a 30-minute cadence and in Sector 33 at a 2-minute cadence. We used light curves produced by the Science Processing Operations Center \citep[SPOC;][]{Jenkins2016}.
All the {\it TESS} data used in this paper can be found in the Mikulski Archive for Space Telescopes (MAST)\footnote{\url{https://mast.stsci.edu/portal/Mashup/Clients/Mast/Portal.html}}: \dataset[10.17909/9jfx-nx45]{http://dx.doi.org/10.17909/9jfx-nx45}. These data are from the Presearch Data Conditioning Simple Aperture Photometry pipeline (PDCSAP; \citet{Stumpe2014, Smith2012}), which strives to remove instrumental systematics while retaining the transits. We normalized and flattened the light curves using the {\tt lightkurve} package in Python \citep{lightkurve2018}. The data were first cleaned by removing outliers with a 10$\sigma$ cutoff threshold. To preserve transit signals during flattening, we applied a transit mask based on known ephemerides. The light curves were then detrended using {\tt lightkurve}’s built-in Savitzky–Golay filter, excluding the masked regions. This process removes low-frequency trends caused by stellar and instrumental variability as well as scattered background light, effectively isolating the astrophysical variability of interest while preserving accurate flux uncertainties.

\subsubsection{KPF Rossiter-McLaughlin Observations} 
The vast majority of existing obliquity measurements have been performed on hot Jupiters, as the signal amplitude scales as $(R_P/ R_\star)^2$. With the newly commissioned Keck Planet Finder \citep[KPF;][]{Gibson2016,Gibson2018,Gibson2020}, we are able to push the limits of stellar obliquity measurements to smaller planets. KPF is a high-resolution, high-stability echelle spectrometer at the W. M. Keck Observatory. It covers a wavelength range of 445–870~nm in two separate channels, with a median resolving power of 98,000. It is designed to characterize exoplanets via Doppler spectroscopy and can achieve a single-measurement precision of 0.3 m/s or better. KPF is wavelength-calibrated using ThAr and UNe lamps and a laser frequency comb from Menlo Systems \citep{Rubenzahl2023}.

The RM spectra of TOI-880 were observed on UTC January 20, 2024, during a transit of TOI-880 c. We obtained 32 exposures with an integration time of 500 seconds. We achieved a typical signal-to-noise ratio (SNR) of 230 at 550~nm. The spectra were reduced with the KPF Data Reduction Pipeline \citep[DRP;][]{Gibson2020}, which is publicly available on \href{https://github.com/Keck-DataReductionPipelines/KPF-Pipeline}{GitHub}. The RV measurements, which were extracted using the CCF method using the {\tt ESPRESSO} masks, are reported in Table \ref{tab:KPF_data}.

\begin{deluxetable}{@{}ccc@{}}
\tablecaption{TOI-880 KPF RM Data \label{tab:KPF_data}}
\tabletypesize{\footnotesize}
\tablehead{\colhead{$\rm Time (BJD-2457000$)} & \colhead{RV (m/s)}  & \colhead{$\rm \sigma_{RV} (m/s)$} }
\startdata
3329.751599 & -0.58 & 0.50 \\ 
3329.757885 & -1.30 & 0.51 \\ 
3329.764233 & -1.14 & 0.49 \\ 
3329.770575 & -2.07 & 0.49 \\ 
3329.776953 & 0.37 & 0.48 \\ 
3329.783238 & 0.45 & 0.48 \\ 
3329.789662 & 2.30 & 0.53 \\ 
3329.798838 & 2.60 & 0.53 \\ 
3329.805155 & 1.80 & 0.50 \\ 
3329.811464 & 1.74 & 0.50 \\ 
3329.817819 & 0.92 & 0.49 \\ 
3329.824116 & 1.01 & 0.48 \\ 
3329.830528 & 0.20 & 0.48 \\ 
3329.836896 & -0.71 & 0.46 \\ 
3329.846241 & -1.24 & 0.47 \\ 
3329.852499 & -1.33 & 0.46 \\ 
3329.858839 & -1.92 & 0.46 \\ 
3329.865154 & -2.28 & 0.47 \\ 
3329.871537 & -1.65 & 0.48 \\ 
3329.881340 & -1.24 & 0.48 \\ 
3329.887795 & 0.73 & 0.46 \\ 
3329.894101 & 0.08 & 0.46 \\ 
3329.900577 & 0.45 & 0.46 \\ 
3329.906797 & 0.53 & 0.46 \\ 
3329.913189 & 1.09 & 0.44 \\ 
3329.919187 & 0.37 & 0.47 \\ 
3329.928332 & 0.38 & 0.45 \\ 
3329.934660 & -0.36 & 0.45 \\ 
3329.941044 & 0.16 & 0.46 \\ 
3329.947407 & -0.20 & 0.45 \\ 
3329.953678 & 0.15 & 0.45 \\ 
3329.960114 & 0.69 & 0.45 \\ \hline
\enddata
\end{deluxetable}

\section{Host Star Spectroscopic Analysis}\label{sec:stellar_para}
The spectroscopic parameters of TOI-880, including $\teff$, log $g$, [Fe/H], and $\vsini$, were derived from a high-resolution, high-SNR, iodine-free spectrum obtained with the High Resolution Echelle Spectrometer on the 10-meter Keck Telescope \citep[Keck/HIRES;][]{Vogt2014}. This observation was taken on UTC October 15, 2021 with an exposure time of 87.1 seconds. It achieved a SNR of about 170 at 550~nm.

We analyzed this spectrum using the {\tt SpecMatch-Syn} pipeline\footnote{\url{https://github.com/petigura/specmatch-syn}} \citep{Petigura2017}, which compares observed spectra to a synthetic spectrum. Details for {\tt SpecMatch-Syn} can be found in \citet{Petigura_thesis}. Briefly, {\tt SpecMatch-Syn} generates a synthetic spectrum by interpolating within a precomputed theoretical grid of spectra with discrete values of $\teff$, [Fe/H], log $g$, and $\vsini$, as outlined in \citet{Coelho2005}. Corrections for stellar rotation and macro-turbulence were applied following the methods in \citet{Hirano2011}. Instrumental broadening was included using a Gaussian with a full width at half maximum (FWHM) of 3.8 HIRES pixels, which matches the typical widths of telluric lines \citep{Petigura_thesis}. The {\tt SpecMatch-Syn} routine then fits the best spectroscopic parameters to five $\sim400$\text{\AA} spectral segments individually and then average the resulting parameters. Known systematic biases of {\tt SpecMatch-Syn} were corrected, such as the bias towards higher log $g$ ($\sim$ 0.1 dex) in earlier-type stars, as seen when compared with the asteroseismic sample of \citet{Huber2013}.

Using Gaia EDR3’s reported parallax of $16.511\pm0.014$ mas \citep{Gaia, EDR3} and the spectroscopic parameters, we derived the stellar parameters. An independent estimate of the stellar radius was obtained via the Stefan-Boltzmann Law, using the spectroscopic $\teff$, the $K$-band magnitude (less affected by extinction than optical magnitudes), and distance information from the Gaia parallax. We used the Python package {\tt Isoclassify} \citep{Huber2017} for isochronal fitting with the models of MESA Isochrones \& Stellar Tracks \citep[MIST,][]{MIST}. The derived stellar parameters for TOI-880 are: $\teff=5050 \pm 100$ K, log $g=4.53 \pm 0.10$, $\rm [Fe/H]=0.23 \pm 0.06$, $\vsini=2.45 \pm 1.0\ km/s$, $\rm M_{\star}=0.87 \pm 0.03\msun$, and $\rm R_{\star}=0.83 \pm 0.03\rsun$.  {\tt Isoclassify} currently does not account for systematic errors across different stellar model grids. As per \citet{Tayar2022}, this may introduce systematic uncertainties of $\sim2\%$ on $\teff$, $\sim4\%$ on $\rm R_{\star}$, and $\sim5\%$ on $\rm M_{\star}$.

SED fitting of TOI-880 using {\tt EXOFASTv2} \citep{Eastman2019} gave a wide range of possible stellar ages: $5.2_{-3.3}^{+4.1}$ Gyr. Also, we could not robustly detect any rotational modulation in the {\it TESS} light curve using the auto-correlation function \citep{McQuillan2014} or the Lomb-Scargle periodogram \citep{Lomb1976,Scargle1982}. Using $\vsini = 1.55 \pm 0.18 \rm ~km/s$ extracted from our RM modeling (described in the next section), the corresponding minimum rotation period from $P=\frac{2 \pi R_\star}{\vsini}$ is $P\sin i_\star = 27.1 \pm 3.3$ days. Based on the gyrochronological relations presented by \citet{Bouma2023}, TOI-880 is likely a mature main-sequence star.

\section{Data Analysis: Joint Light Curve and RM Fit} \label{sec:analysis}
We used the {\tt allesfitter} package \citep{allesfitter-paper, allesfitter-code} to determine the sky-projected spin-orbit angle, $\lambda$, by jointly modeling the {\it TESS} light curves (sector 6 and 33) and the RM data for TOI-880 c from KPF. The {\it TESS} light curves include 6 transits of TOI-880 c, 18 transits of TOI-880 b, and 4 transits of TOI-880 d. Our model accounted for the different cadence of {\it TESS} light curves. For each planet, the transit parameters include: $R_p / R_{\star}$ (radius ratio), $(R_{\star} + R_p) / a$ (the sum of stellar and companion radii, divided by the semi-major axis), $\cos{i_p}$ (the cosine of the orbital inclination), $T_0$ (the transit mid-time), $P$ (the planet's orbital period).  Since we do not have orbital radial velocity data, we did not fit for the RV semi-amplitudes, K. Similarly, we assumed circular orbits for all the planets. In addition to the RM effect, our model for the RV time series included an additive {\it ad hoc} quadratic function of time, to account for orbital motion as well as stellar activity. The RM effect introduced two additional parameters $\lambda_c$ (the sky-projected stellar obliquity using planet c) and $\vsini$ (the projected stellar rotation velocity). We adopted a quadratic limb darkening law specified by $q_{1; \mathrm{{\it TESS}}}$, $q_{2; \mathrm{{\it TESS}}}$, $q_{1; \mathrm{KPF}}$, and $q_{2; \mathrm{KPF}}$, where $q_1$ and $q_2$ refer to the parameterization advocated by \citep{Kipping2013}. The errors (white noise) per instrument are described by two free parameters: $\ln{\sigma_\mathrm{{\it TESS}}}$, which is the error scaling for TESS, and $\ln{\sigma_\mathrm{jitter; KPF}}$, which represents the jitter term added in quadrature to the KPF uncertainties.

\begin{figure}[htb]
    \centering 
    \includegraphics[width=0.47\textwidth]{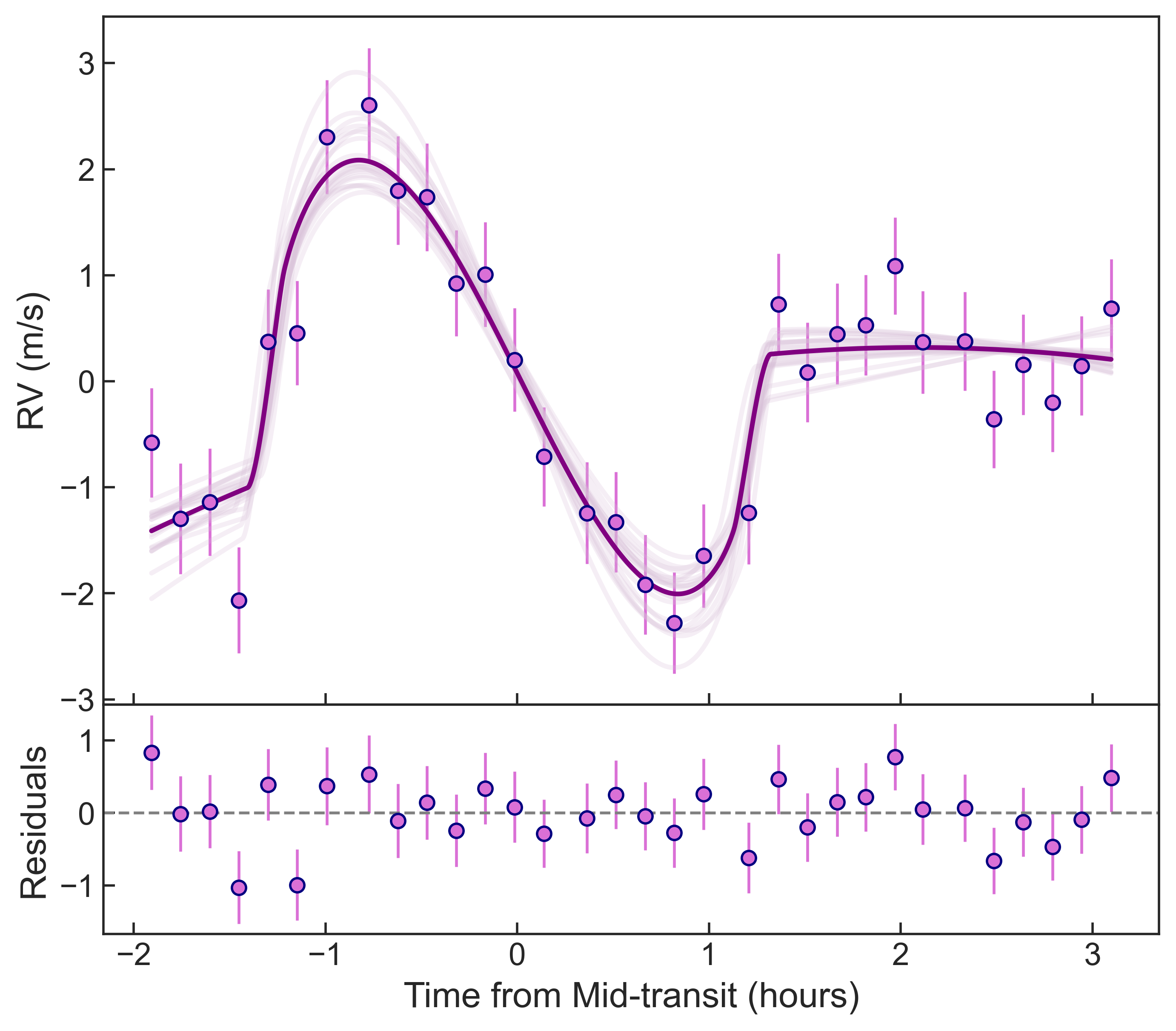}
    \caption{\normalsize The radial velocity variations during the transit of TOI-880 c on UTC January 20, 2024, as observed by KPF. In the top panel, the fainter pink curves are from random posterior samples. The darker purple line is the best-fit model with a sky-projected obliquity of $\lambda_c = -7.4_{-7.2}^{+6.8}$$^{\circ}$. The bottom panel shows the residuals, subtracting the best-fit model.}
\label{Fig:b_RM}
\end{figure}

\begin{figure*}[htbp]
    \centering
    \includegraphics[width=0.31\textwidth]{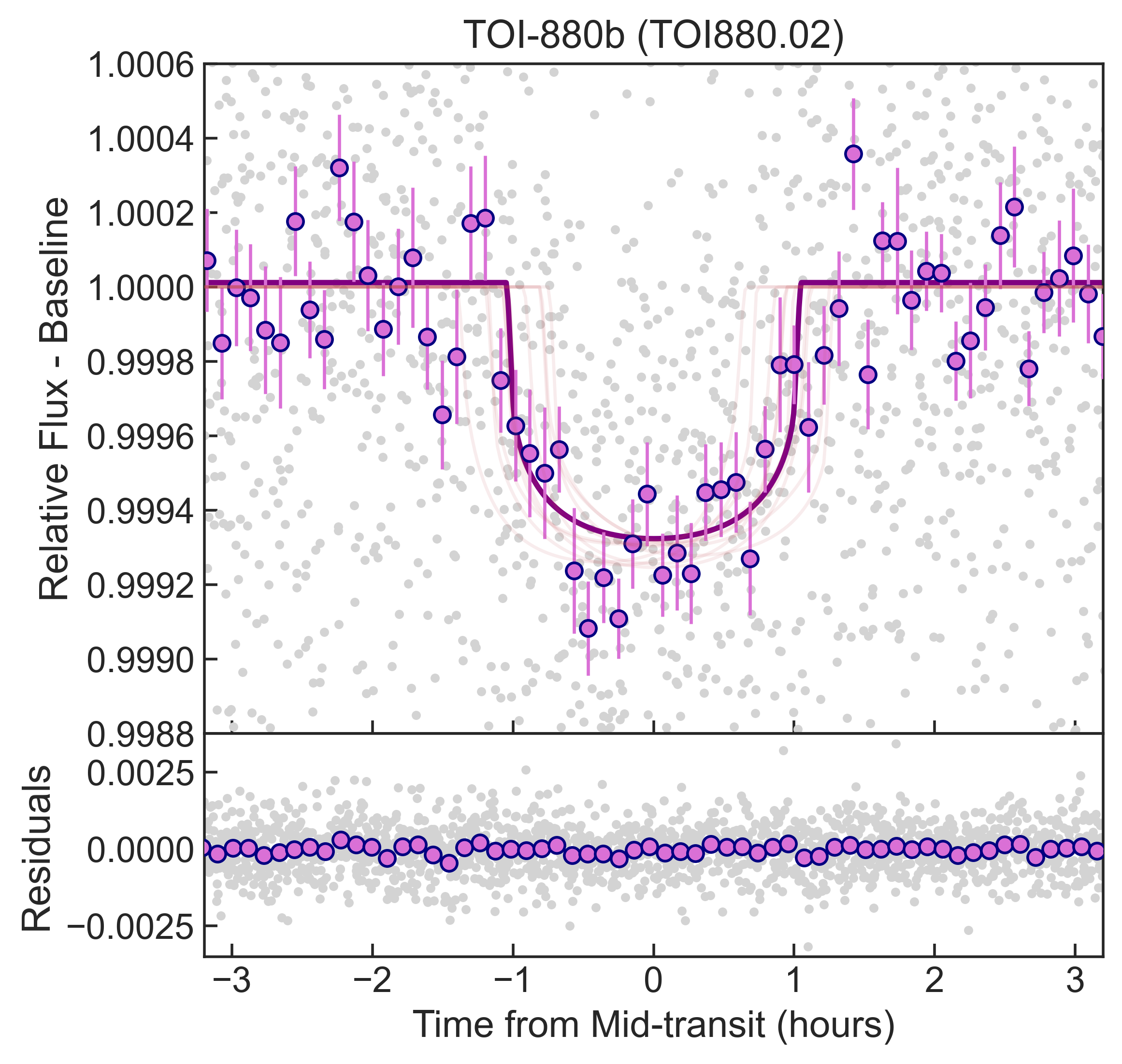}
    \includegraphics[width=0.31\textwidth]{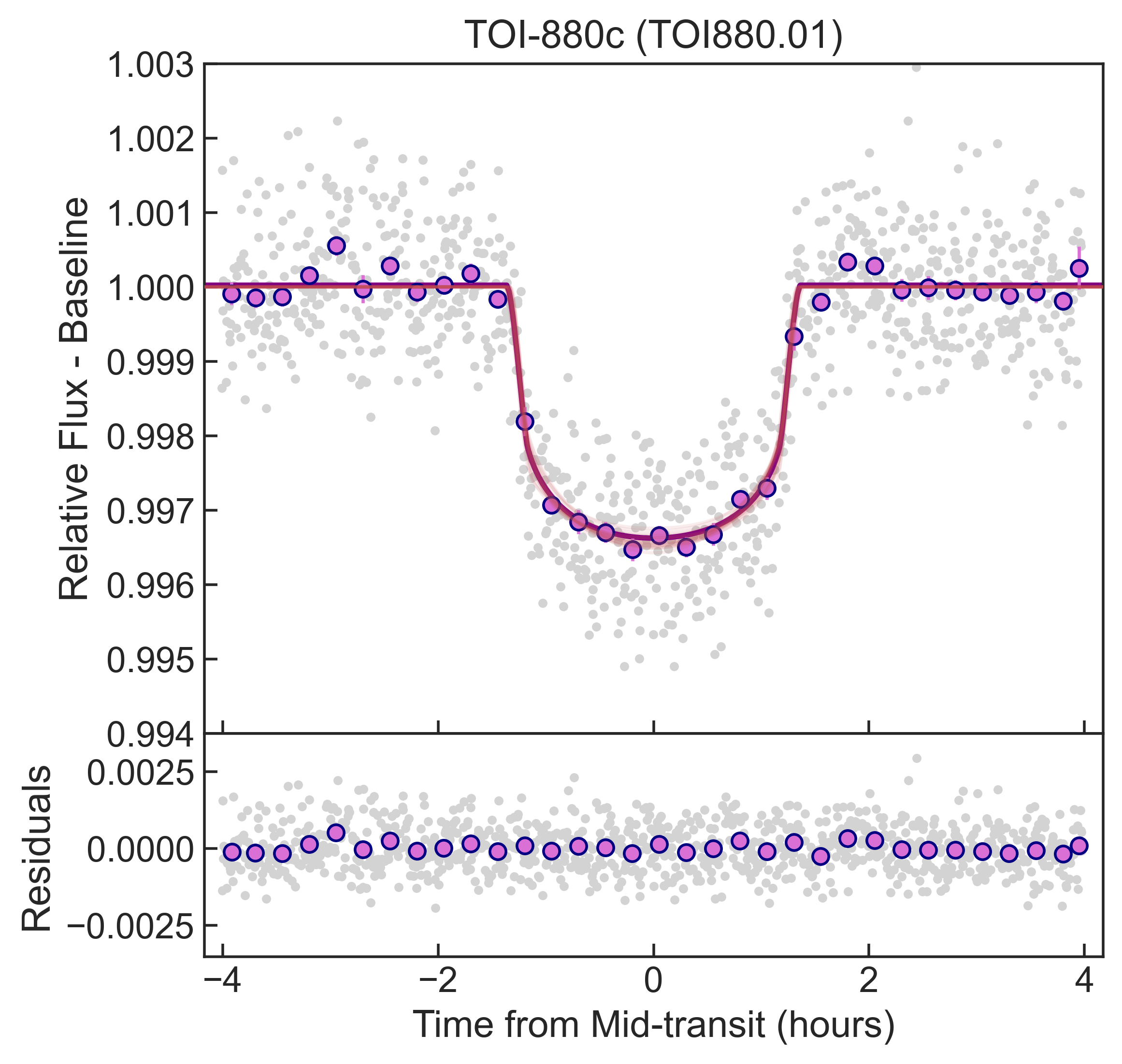}
    \includegraphics[width=0.31\textwidth]{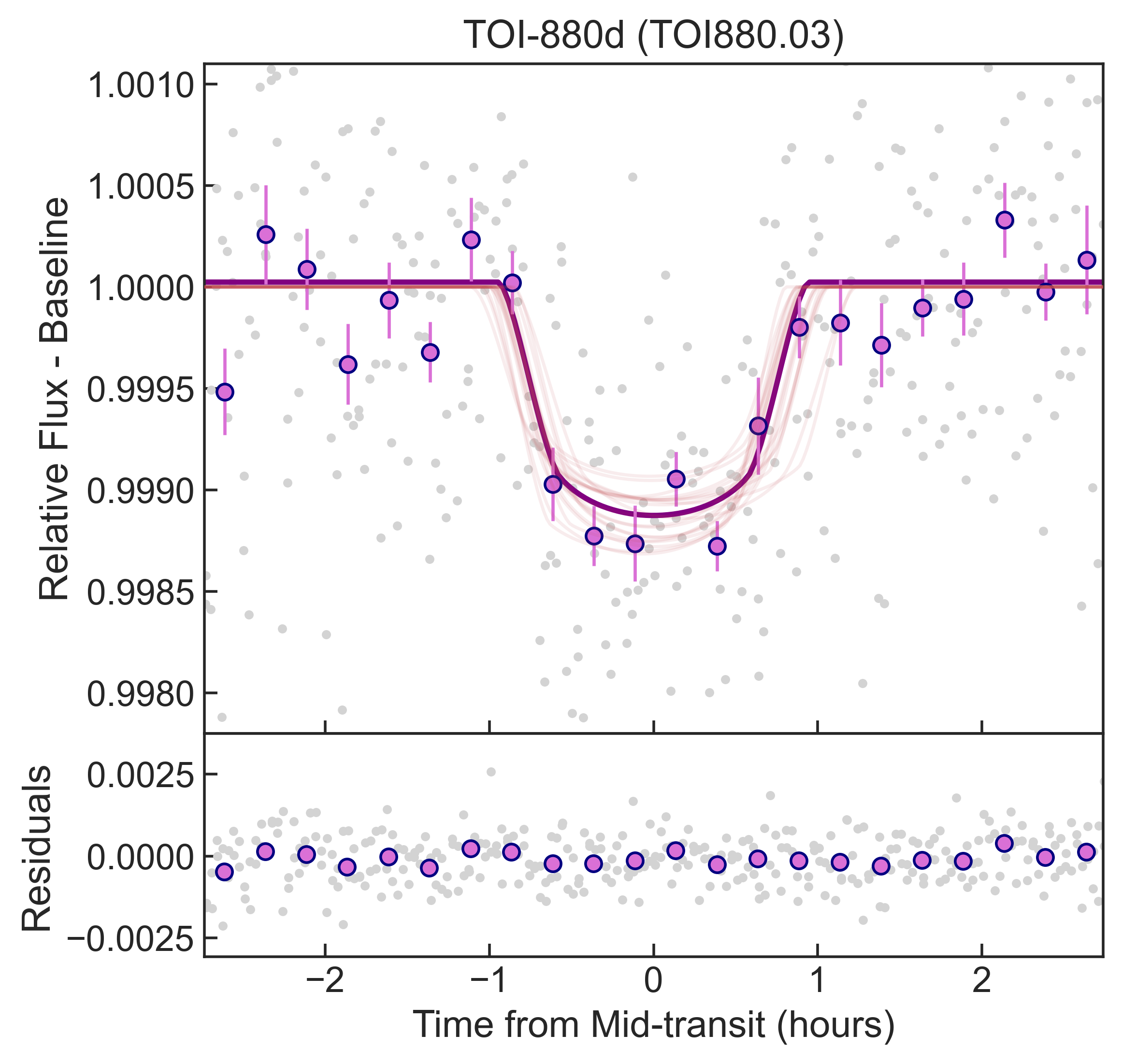}
    \caption{\normalsize Normalized flux verses time from mid-transit time (in hours) for TOI-880 b (left), TOI-880 c (middle), TOI-880 d (right). The planets are ordered in terms of increasing orbital periods. The star was observed at 2-minute cadence in sector 33. The pink points with blue edges are the phase-folded, binned data. The faint red lines are 20 curves from random posterior samples, and the darker purple line is the median fitted model from {\tt allesfitter}. The bottom panels show the residuals, subtracting the best-fit model.}
\label{Fig:TESS_transit}
\end{figure*}

We set uniform priors $\mathcal{U}$(a, b) on our input parameters. We also added a Gaussian prior on the stellar mean density using the spectroscopic results from {\tt Isoclassify}. We employed {\tt allesfitter} to sample the posterior distribution with the dynamic Nested Sampling algorithm \citep{Higson2019}. We used 500 live points at the beginning of a Nested Sampling run. The live points were updated using a random walk sampling method. The median and 1-$\sigma$ confidence interval of the posterior distributions are reported in Table \ref{tab:fit_para}. The best-fit RM model for planet c and the {\it TESS} light curves for planets b, c, and d are shown in Figure \ref{Fig:b_RM} and Figure \ref{Fig:TESS_transit}, respectively. Parameters that were derived from the combination of transit and stellar parameters are shown in Table \ref{tab:deri_para}.

\begin{deluxetable*}{@{}cccc@{}}
\tablecaption{{\tt allesfitter} Fitted Parameters\label{tab:fit_para}}
\tabletypesize{\footnotesize}
\tablehead{ \colhead{parameter} & \colhead{priors}  & \colhead{value} & \colhead{unit} }
\startdata
$R_b / R_\star$ & $\mathcal{U}$(0, 0.1) & $0.02417\pm0.00090$ &  \\ 
$(R_\star + R_b) / a_b$ & $\mathcal{U}$(0, 0.5) & $0.1131_{-0.0042}^{+0.0048}$ & \\ 
$\cos{i_b}$ & $\mathcal{U}$(0, 0.2) & $0.038_{-0.022}^{+0.016}$ &  \\ 
$T_{0;b}-2457000^{\ast}$ & $\mathcal{U}$(2202.195333, 2202.395333) & $2400.6128_{-0.0093}^{+0.020}$ & $\mathrm{BJD}$ \\ 
$P_b$ & $\mathcal{U}$(2.473594, 2.673594) & $2.57571_{-0.00012}^{+0.00028}$ & $\mathrm{day}$ \\ 
$\sqrt{e_b} \cos{\omega_b}$ & $0.0$ & fixed &  \\ 
$\sqrt{e_b} \sin{\omega_b}$ & $0.0$ & fixed &  \\ \hline
$R_c / R_\star$ & $\mathcal{U}$(0, 0.1) & $0.05467\pm0.00095$ &  \\
$(R_\star + R_c) / a_c$ & $\mathcal{U}$(0, 0.5) & $0.0640_{-0.0025}^{+0.0031}$ &  \\ 
$\cos{i_c}$ & $\mathcal{U}$(0, 0.2) & $0.0314_{-0.0052}^{+0.0056}$ & \\ 
$T_{0;c}-2457000^{\ast}$ & $\mathcal{U}$(2224.730978, 2224.930978) & $2403.67511\pm0.00040$ & $\mathrm{BJD}$ \\ 
$P_c$ & $\mathcal{U}$(6.28727889,6.48727889) & $6.3872703_{-0.0000073}^{+0.0000069}$ & $\mathrm{day}$ \\ 
$\sqrt{e_c} \cos{\omega_c}$ & $0.0$ & fixed &  \\ 
$\sqrt{e_c} \sin{\omega_c}$ & $0.0$ & fixed &  \\ \hline
$R_d / R_\star$ & $\mathcal{U}$(0, 0.1) & $0.0376\pm0.0020$ &  \\ 
$(R_\star + R_d) / a_d$ & $\mathcal{U}$(0, 0.5) & $0.0364_{-0.0014}^{+0.0016}$ &  \\ 
$\cos{i_d}$ & $\mathcal{U}$(0, 0.2) & $0.0321_{-0.0016}^{+0.0017}$ & \\ 
$T_{0;d}-2457000^{\ast}$ & $\mathcal{U}$(2220.559691, 2220.759691) & $2406.9736_{-0.0025}^{+0.0027}$ & $\mathrm{BJD}$ \\ 
$P_d$ & $\mathcal{U}$(14.231885, 14.431885) & $14.33201\pm0.00011$ & $\mathrm{day}$ \\ 
$\sqrt{e_d} \cos{\omega_d}$ & $0.0$ & fixed &  \\ 
$\sqrt{e_d} \sin{\omega_d}$ & $0.0$ & fixed &  \\ \hline
$q_{1; \mathrm{{\it TESS}6}}$ & $\mathcal{U}$(0, 1) & $0.42_{-0.17}^{+0.22}$ & \\ 
$q_{2; \mathrm{{\it TESS}6}}$ & $\mathcal{U}$(0, 1) & $0.51\pm0.32$ &  \\ 
$q_{1; \mathrm{{\it TESS}33}}$ & $\mathcal{U}$(0, 1) & $0.47_{-0.18}^{+0.24}$ & \\ 
$q_{2; \mathrm{{\it TESS}33}}$ & $\mathcal{U}$(0, 1) & $0.25_{-0.17}^{+0.28}$ &  \\ 
$q_{1; \mathrm{KPF}}$ & $\mathcal{U}$(0, 1) & $0.77_{-0.18}^{+0.14}$ &  \\ 
$q_{2; \mathrm{KPF}}$ & $\mathcal{U}$(0, 1) & $0.76_{-0.27}^{+0.17}$ &  \\ 
$\ln{\sigma_\mathrm{{\it TESS}6}}$ & $\mathcal{U}$(-10, 0) & $-7.949\pm0.050$ & $\ln{ \mathrm{rel. flux.} }$ \\ 
$\ln{\sigma_\mathrm{{\it TESS}33}}$ & $\mathcal{U}$(-10, 0) & $-7.178\pm0.012$ & $\ln{ \mathrm{rel. flux.} }$ \\ 
$\ln{\sigma_\mathrm{jitter; KPF}}$ & $\mathcal{U}$(-10, 0) & $-9.15_{-0.58}^{+0.66}$ & $\ln{ \mathrm{km/s} }$ \\ 
$\lambda_c$ & $\mathcal{U}$(-180, 180) & $-7.4_{-7.2}^{+6.8}$ & degree \\ 
$vsini_{\star}$ & $\mathcal{U}$(0, 10) & $1.57_{-0.16}^{+0.18}$ & $\mathrm{km/s}$ \\ 
\enddata
\tablecomments{$\mathcal{U}$(a, b) means an uniform prior with range (a, b). \\
The subscript `{\it TESS}6' means {\it TESS} sector 6 and `{\it TESS}33' for {\it TESS} sector 33. \\
$^{\ast}${\tt allesfitter} will automatically shift the input prior epoch into the data center. For TOI-880 c, the epoch is shifted by 28 periods, 77 periods for TOI-880 b, and 13 periods for TOI-880 d. }
\end{deluxetable*}

\begin{deluxetable*}{@{}cc@{}}
\tablecaption{{\tt allesfitter} Derived Parameters\label{tab:deri_para}}
\tabletypesize{\footnotesize}
\tablehead{ \colhead{parameter} & \colhead{value} }
\startdata
Semi-major axis b over host radius; $a_\mathrm{b}/R_\star$ & $9.05_{-0.37}^{+0.35}$   \\ 
Companion radius b; $R_\mathrm{b}$ ($\mathrm{R_{\oplus}}$) & $2.19\pm0.11$   \\ 
Semi-major axis b; $a_\mathrm{b}$ (AU) & $0.0349\pm0.0019$   \\ 
Inclination b; $i_\mathrm{b}$ (deg) & $87.80_{-0.95}^{+1.3}$   \\ 
Impact parameter b; $b_\mathrm{tra;b}$ & $0.35_{-0.20}^{+0.14}$   \\ 
Full-transit duration b; $T_\mathrm{full;b}$ (h) & $1.97_{-0.12}^{+0.13}$   \\ 
Equilibrium temperature b; $T_\mathrm{eq;b}$ (K) & $1085_{-29}^{+31}$   \\ 
Transit depth b; $\delta_\mathrm{tr; b; {\it TESS}6}$ (ppt) & $0.706_{-0.053}^{+0.066}$   \\ 
Transit depth b; $\delta_\mathrm{tr; b; {\it TESS}33}$ (ppt) & $0.688_{-0.041}^{+0.045}$   \\ \hline
Semi-major axis c over host radius; $a_\mathrm{c}/R_\star$ & $16.48_{-0.76}^{+0.67}$   \\ 
Companion radius c; $R_\mathrm{c}$ ($\mathrm{R_{\oplus}}$) & $4.95\pm0.20$   \\ 
Semi-major axis c; $a_\mathrm{c}$ (AU) & $0.0635\pm0.0036$   \\ 
Inclination c; $i_\mathrm{c}$ (deg) & $88.20_{-0.32}^{+0.30}$   \\ 
Impact parameter c; $b_\mathrm{tra;c}$ & $0.517_{-0.070}^{+0.064}$   \\ 
Full-transit duration c; $T_\mathrm{full;c}$ (h) & $2.343\pm0.045$   \\ 
Equilibrium temperature c; $T_\mathrm{eq;c}$ (K) & $805_{-23}^{+25}$   \\ 
Transit depth c; $\delta_\mathrm{tr; c; {\it TESS}6}$ (ppt) & $3.43_{-0.12}^{+0.12}$   \\ 
Transit depth c; $\delta_\mathrm{tr; c; {\it TESS}33}$ (ppt) & $3.405\pm0.063$   \\ \hline
Semi-major axis d over host radius; $a_\mathrm{d}/R_\star$ & $28.5_{-1.2}^{+1.1}$   \\ 
Companion radius d; $R_\mathrm{d}$ ($\mathrm{R_{\oplus}}$) & $3.40_{-0.21}^{+0.22}$   \\ 
Semi-major axis d; $a_\mathrm{d}$ (AU) & $0.1098\pm0.0060$   \\ 
Inclination d; $i_\mathrm{d}$ (deg) & $88.161_{-0.100}^{+0.090}$   \\ 
Impact parameter d; $b_\mathrm{tra;d}$ & $0.915_{-0.016}^{+0.013}$   \\ 
Full-transit duration d; $T_\mathrm{full;d}$ (h) & $1.14\pm0.16$   \\ 
Equilibrium temperature d; $T_\mathrm{eq;d}$ (K) & $612_{-17}^{+18}$   \\ 
Transit depth d; $\delta_\mathrm{tr; d; {\it TESS}6}$ (ppt) & $1.11\pm0.11$   \\ 
Transit depth d; $\delta_\mathrm{tr; d; {\it TESS}33}$ (ppt) & $1.138_{-0.085}^{+0.11}$   \\ \hline
Period ratio; $P_\mathrm{c} / P_\mathrm{b}$ & $2.47981_{-0.00027}^{+0.00012}$   \\
Period ratio; $P_\mathrm{d} / P_\mathrm{b}$ & $5.56429_{-0.00060}^{+0.00027}$   \\ 
Period ratio; $P_\mathrm{d} / P_\mathrm{c}$ & $2.243840_{-0.000016}^{+0.000017}$   \\ 
Limb darkening; $u_\mathrm{1; {\it TESS}6}$ & $0.61\pm0.37$   \\ 
Limb darkening; $u_\mathrm{2; {\it TESS}6}$ & $-0.01_{-0.35}^{+0.45}$   \\ 
Limb darkening; $u_\mathrm{1; {\it TESS}33}$ & $0.35_{-0.22}^{+0.25}$   \\ 
Limb darkening; $u_\mathrm{2; {\it TESS}33}$ & $0.34_{-0.38}^{+0.33}$   \\ 
Limb darkening; $u_\mathrm{1; KPF}$ & $1.27_{-0.44}^{+0.30}$   \\ 
Limb darkening; $u_\mathrm{2; KPF}$ & $-0.43_{-0.27}^{+0.46}$   \\ 
Combined host density from all orbits; $\rho_\mathrm{\star; combined}$ (cgs) & $2.11\pm0.26$   \\ 
\enddata
\tablecomments{TOI-880 b (TOI-880.02) is denoted as companion subscript b, TOI-880 c (TOI-880.01) as c,  and TOI-880 d (TOI-880.03) as d. \\
The subscript `{\it TESS}6' means {\it TESS} sector 6 and `{\it TESS}33' for {\it TESS} sector 33.}
\end{deluxetable*}

\section{Coplanarity of the TOI-880 System}
\label{sec:mul-tran}

The fact that there are three transiting planets in the TOI-880 system suggests a coplanar configuration. Our transit modeling yields orbital inclinations of $87.80_{-0.95}^{+1.30\circ}$, $88.20_{-0.32}^{+0.30\circ}$, and $88.16_{-0.10}^{+0.09\circ}$, for planet b c d respectively. Assuming a common longitude of ascending node, the mutual inclinations among the TOI-880 planets appear to $\lesssim 2^\circ$. This is consistent with the results of \citet{Fabrycky2014} who found statistical evidence for low mutual inclinations (1-2$^\circ$) in typical
Kepler multi-planet systems.

A planetary transit is observable only if the line of sight lies within a narrow strip of angular width $\sim R_\star/a$ centered on the planet’s orbital plane. For multiple planets to be observed in transit, these transit strips must overlap on the unit sphere, and the observer must be located within this overlapping region. As the mutual inclinations between planets increase, the area of overlap diminishes, significantly reducing the likelihood of detecting multiple transiting planets from a random vantage point. We leverage this effect to further quantify the coplanarity of the TOI-880 planets.

We conducted Monte Carlo simulations following the methodology of \citet{Lissauer2011}. We fixed the orbital periods of TOI-880 planets and assumed circular orbits. The mutual inclinations were modeled using a Rayleigh distribution $f(\Delta I, \sigma) = \frac{\Delta I}{\sigma^2} \exp{-\frac{\Delta I^2}{2\sigma^2}}$ where $\sigma$ is the only parameter of the distribution with mean $\sqrt{\frac{\pi}{2}}\sigma$ and the standard deviation $\sqrt{\frac{4-\pi}{2}}\sigma$. We generated a grid of $\sigma$ from $0^{\circ}$ to $30^{\circ}$ in increments of $0.3^{\circ}$. At each grid point, we simulated 1000 systems to calculate the probability that a randomly oriented observer could see all three planets transit. As expected, this probability declines quickly as the mutual inclination increases. We estimate that there is only a 1\% chance that {\it TESS} has the right vantage point so as to see all three TOI-880 planets transit if the system has a mean mutual inclination $7^{\circ}$ or more. Admittedly, if the intrinsic planet multiplicity is higher than 3, the constraint on the mutual inclination becomes less stringent. To recap, the existing observations for TOI-880 favor a coplanar configuration for the three transiting planets. The stellar obliquity measured for TOI-880 c is likely representative of the other two planets.

\section{Nodal Precession of the TOI-880 Planets}
\label{sec:precession}

\begin{deluxetable*}{ccccc}[ht]
\tablecaption{The eigensystem of the precession matrix $\boldsymbol{M}$ in Equation \ref{eqn:xi} specified by the parameters of Table \ref{tab:deri_para}. } \label{tab:eigen}
\tablehead{
Eigenmode & No. 1 & No. 2  & No. 3 & Planet }
\startdata
Eigenvalues $(\rm{rad}\,\rm{yr}^{-1})$ & $2.0 \times 10^{-2}$  & $9.4 \times 10^{-3}$ & $3.0 \times 10^{-6}$ & \\
\hline
  & $-9.6 \times 10^{-1}$  & $-6.5 \times 10^{-1}$ & $5.8 \times 10^{-1}$& b \\
Eigenvectors  & $2.5 \times 10^{-1}$  & $-3.3 \times 10^{-1}$ & $5.8 \times 10^{-1}$& c\\
& $-8.7 \times 10^{-2}$  & $6.9 \times 10^{-1}$ & $5.8 \times 10^{-1}$& d \\
\hline
Constant $c_{k}$ & $0.063 \times \exp{(1.09\,\iota)}$  & $0.151 \times \exp{(-1.04\,\iota)}$ & $0.324 \times \exp{(1.27\,\iota)} $ &\\
\enddata
\tablecomments{The constants in the last rwo were determined for the initial condition where planets are perfect coplanar but mildly misaigned with the host star true stellar obliquities $I_{\mathrm{b}} = I_{\mathrm{c}} = I_{\mathrm{d}}= 5^{\circ}$ and ascending nodes $\Omega_{\rm{b}} = \Omega_{\rm{c}} = \Omega_{\rm{d}} = 90^{\circ}$.}
\end{deluxetable*}

\begin{figure*}[htbp]
    \centering
    \includegraphics[width=0.6\textwidth]{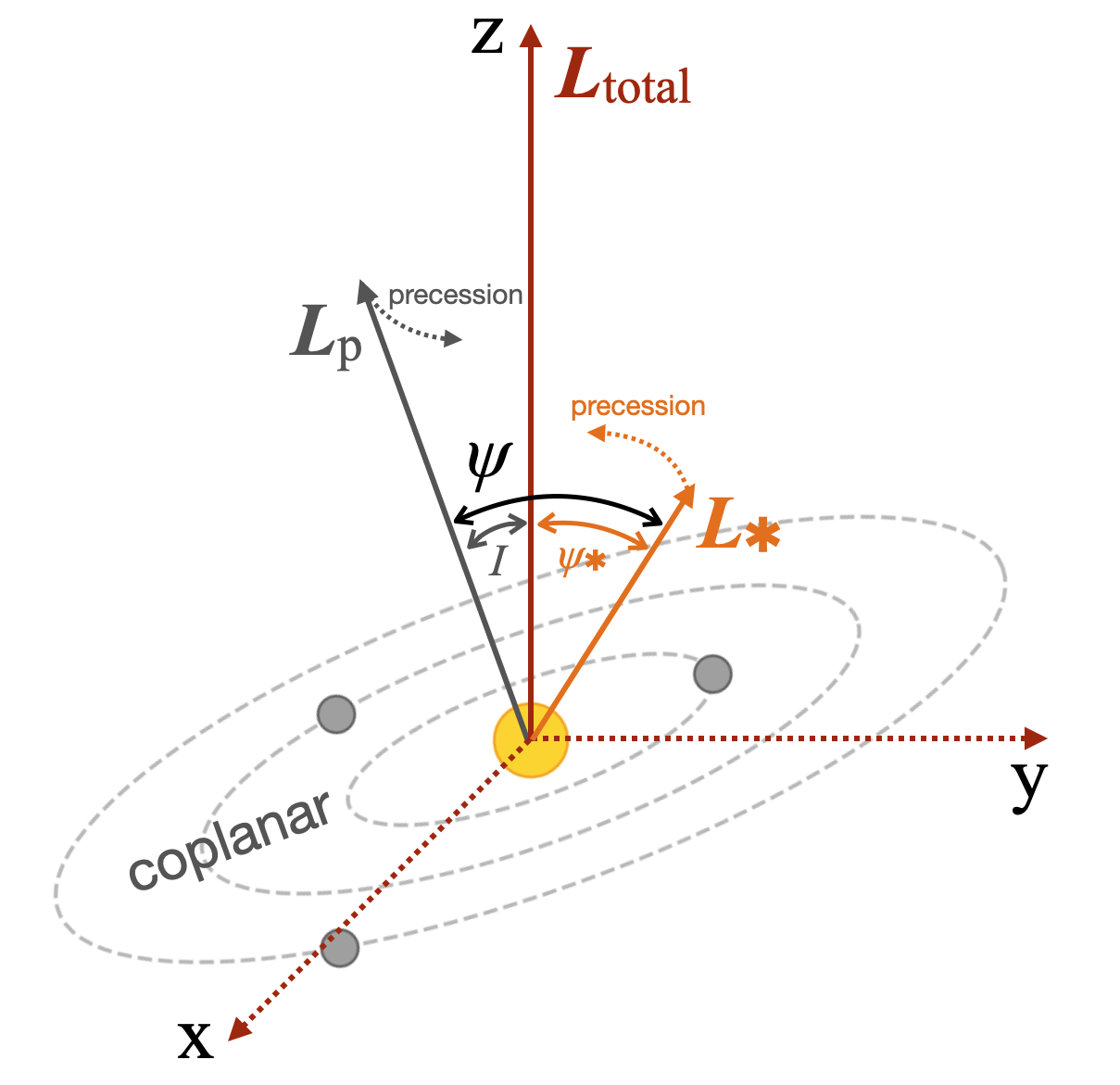}
    \caption{Precession geometry of the TOI-880 system. In this framework, the z-axis is fixed to the system’s total angular momentum  ($L_{\mathrm{total}}$)  --- the sum of the stellar rotational component ($L_{\star}$) and the total planetary orbital component ($L_{\mathrm{p}}$). Specifically for TOI-880, the system has $L_{\star} \sim L_{\mathrm{p}}$ with both components being non-negligible and simultaneously precessing around the system’s total angular momentum. Per Laplace-Lagrange theory, the planetary inclinations in the TOI-880 system undergo tiny oscillations around a common orbital plane ($I = I_{\mathrm{b}} = I_{\mathrm{c}} = I_{\mathrm{d}}$), and thus the total orbital angular momentum is $L_{\mathrm{p}}=L_{\mathrm{p,b}} + L_{\mathrm{p,c}} + L_{\mathrm{p,d}}$. Hence, the true stellar obliquity is simply derived from $\psi = I + \psi_{\star}$.}
    \label{Fig:dyn_framework}
    \vspace{1cm} 
    \includegraphics[width=0.85\textwidth]{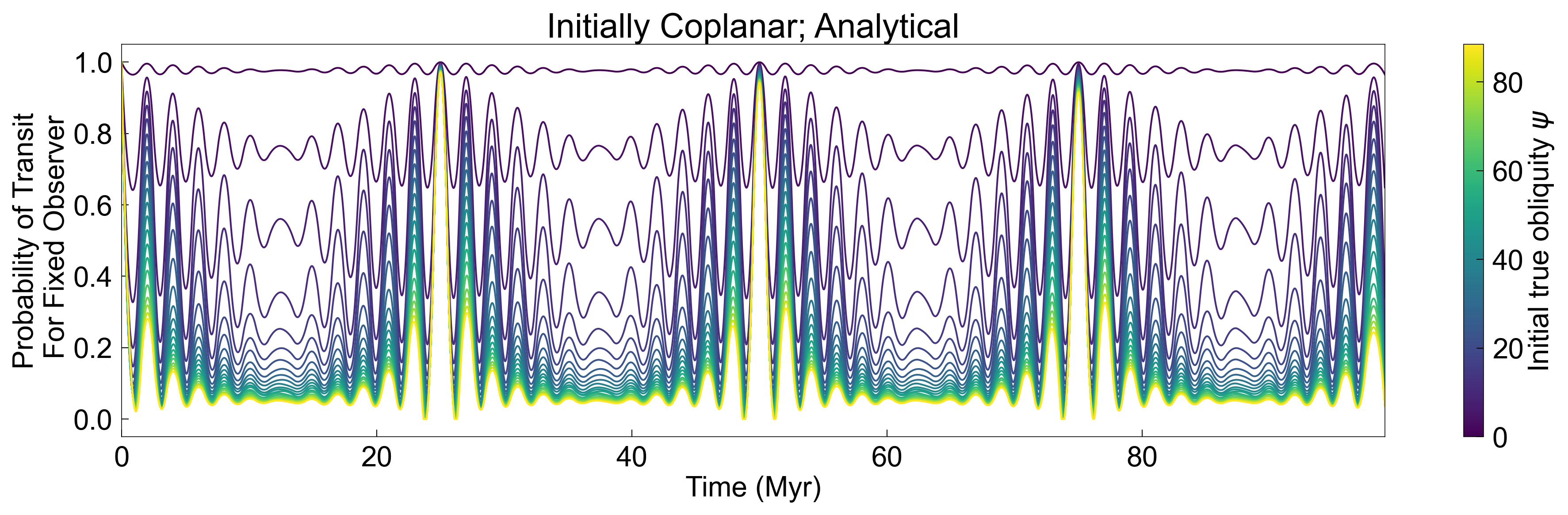}
    \caption{\normalsize Time-varying probability that a fixed observer sees all three TOI-880 planets transit. The color of the solid lines indicate the initial true obliquity of the common orbital plane. As the planets undergo nodal precession, a fixed observer will no longer see all three planets transit. However, because of the low $J_2$, the timescale if of order Myr hence unlikely to be observable for TOI-880.}
    \label{Fig:dyn_prob_mt_fixedob}
\end{figure*}

Even if the TOI-880 planets were initially co-planar,  a non-zero stellar obliquity coupled with differential nodal precession induced by the stellar quadrupole moment may still excite mutual inclinations between the planets \citep{Spalding2016}.  To check on this possibility, we ran a differential nodal precession model similar to that presented in \citet{Barnes2013} and \citet{Spalding2016} (see also, \citealt{Teng2025}). The $z$ axis of our simulation is defined by the total angular momentum of the system $L_{\rm total}$, i.e. the sum of angular momenta of the planets $L_{\mathrm{p,b}},\ L_{\mathrm{p,c}},\ L_{\mathrm{p,d}}$ and the angular momentum of the host star's rotation $L_\star$ (Figure \ref{Fig:dyn_framework}). In the case of TOI-880, the stellar rotational angular momentum is $L_{\star} = 0.93 \times 10^{41}~\text{kg m}^2\text{s}^{-1}$ where we assumed a 27-day stellar rotation period (inferred from rotational broadening), and a momentum of inertia coefficient of $\mathbb{C}=$0.059 similar to the sun ($L_\star = \mathbb{C}M_\star R_\star^2 \omega$).

On the other hand, the planets' orbital angular momenta are $L_{\mathrm{p,b}} = 0.27 \times 10^{41}~\text{kg m}^2\text{s}^{-1}$, $L_{\mathrm{p,c}} = 1.35 \times 10^{41}~\text{kg m}^2\text{s}^{-1}$, and $L_{\mathrm{p,d}} = 0.91 \times 10^{41}~\text{kg m}^2\text{s}^{-1}$ assuming zero eccentricities. In these calculations, we used the mass-radius relationship of \citet{Chen2017} and obtained a median mass of $5.8\,M_{\oplus}$ for TOI-880 b, $21.5\,M_{\oplus}$ for TOI-880 c, and $11.1\,M_{\oplus}$ for TOI-880 d. Nielsen et al. (priv. comm.) obtained RV data for TOI-880, and measured a mass of $3.1\pm0.6\, M_{\oplus}$ for planet b, $22.8\pm0.8\,M_{\oplus}$ for planet c, and $13.1\pm1.2\,M_{\oplus}$ for planet d. These measured values are broadly consistent with the masses we adopted from the mass-radius relationship \citet{Chen2017}.

$L_{\rm total}$ remains fixed according to an external inertial observer. We denote the angle between $z$ axis ($L_{\rm total}$) and the angular momentum of each planet e.g. $I_{b,c,d}$. This is not to be confused with the orbital inclination as viewed from the observer with $i_{b,c,d}$ presented in the transit modeling.  We initialized the three TOI-880 planets in the same orbital plane, i.e., $I_{\rm{b}}=I_{\rm{c}}=I_{\rm{d}}$ and $\Omega_{\rm{b}}=\Omega_{\rm{c}}=\Omega_{\rm{d}}$. In other words, we assumed that they were born out of the same thin protoplanetary disk.

We now calculate the secular coupling between the planets. Since the planets are far from resonances, we employed the Laplace-Lagrange formulation:
\begin{equation}\label{eqn:xi}
    \frac{\mathrm{d}\boldsymbol{\xi}}{\mathrm{d}t} = -\iota\,\boldsymbol{M} \cdot \boldsymbol{\xi},
\end{equation}
where $\boldsymbol{\xi}$ is the complex inclination vector, defined by $\xi_{i} = I_{i} \cos\Omega_{i} + \iota\,I_{i}\sin\Omega_{i}$, $\iota$ is imaginary number, and $\boldsymbol{M}$ is a matrix describing the nodal precession rates of planets. The expansion of matrix $\boldsymbol{M}$ can be found in \citet{Spalding2016}. Since, the $L_{\mathrm{p}}$ and  $L_\star$ are comparable for TOI-880, the full evolution of the system is the precession of both $L_{\rm p}$ and $L_{\star}$ around the total angular momentum vector $L_{\rm total}$ as pointed out by \citet{Barnes2013}. We modified the nodal precession rate of planets in Eqn \ref{eqn:xi} with Eqn 6 and 9 of \citet{Barnes2013}. We adopted a low $J_2 = 10^{-6}$ consistent with the absence of rotational modulation in the {\it TESS} light curve and the low projected rotational velocity ($\rm v\sin i$, which implies a 27-day rotation). For reference, the solar $J_2$ is $2 \times 10^{-7}$ \citep{Pireaux}.

The general solution of the differential equation (Equation \ref{eqn:xi}) is the sum of normal modes, and it can be written in terms of eigenvalues, $\lambda_{k}$, and eigenvectors, $\boldsymbol{\xi}_{k}$, as:
\begin{equation}\label{eqn:sol}
    \boldsymbol{\xi}(t) = \sum_{k}^{} c_{k}  \exp(-\iota \lambda_{k} t) \boldsymbol{\xi}_{k},
\end{equation}
where the complex constants, $c_{k}$, can be determined from the initial conditions. We found the eigenfrequencies (precession rates) to be $2.0\times 10^{-2}\,\rm{rad\,yr}^{-1}$ and $9.4\times 10^{-3}\,\rm{rad\,yr}^{-1}$, and $3.0\times 10^{-6}\,\rm{rad\,yr}^{-1}$. See Table \ref{tab:eigen} for the corresponding eigenvectors.  In other words, the secular forcing between the planets (first two eigen frequencies) is stronger than the precession due to the stellar quadrupole (last eigenfrequency). We note that even the angular momenta $L_{\mathrm{p}}$ and  $L_\star$ are comparable, the relatively slow rotation of TOI-880 generates only a weak stellar quadrupole moment ($J_2$) and thus a weak torque on the planet. Qualitatively, the three TOI-880 planets are well-coupled into a single plane $L_{\mathrm{p}}$ by secular forces, while the  $L_{\mathrm{p}}$ and  $L_\star$ both precess slowly around $L_{\rm total}$ slowly. This is illustrated in Fig. \ref{Fig:dyn_framework}.

More quantitatively, the true stellar obliquity $\Psi$ of $L_{\mathrm{p}}$ to the stellar rotation axis $L_\star$ is:
\begin{equation}
    \psi = I + \psi_{\star},
\end{equation}
with 
\begin{equation}
    \frac{\sin I}{\sin \psi_{\star}} = \frac{L_{\star}}{L_{\mathrm{p}}},
\end{equation}
where $\psi_{\star}$ is the angle between the system's total angular momentum and stellar rotational angular momentum. 

If the common orbital plane of the TOI-880 planets has a non-zero stellar obliquity as predicted by several theoretical models that tilt whole planetary systems \citep{Batygin2012,Lai2011}, the nodal precession due to the stellar quadrupole can eventually cause the planets to precess out of a transiting configuration for a fixed observer. Figure \ref{Fig:dyn_prob_mt_fixedob} shows the probability that all three planets remain transiting as a function of the initial stellar obliquity for a fixed observer, computed using the method in Section \ref{sec:mul-tran}. If the planets are initially well-aligned with the host star, they can remain in a transiting configuration almost indefinitely (blue curves in Figure \ref{Fig:dyn_prob_mt_fixedob}). 
However, if the stellar obliquity is high enough ($\geq 20^{\circ}$)
the planets will cease to transit on a timescale of Myr comparable to last eigenperiod in Table \ref{tab:eigen}. In summary, even though our analysis reveals that the planets are strongly coupled with each other by secular effect and hence resistant to differential nodal precession due to the stellar quadrupole.  The overall precession of $L_{\mathrm{p}}$  is too slow to produce detectable transit duration variations on decade-long timescales.

\section{Atmospheric Mass Loss}
\label{sec:atm}
Exoplanets can lose atmospheric mass through mechanisms such as photoevaporation \citep[e.g.,][]{OwenWu2017}, where high-energy radiation from the host star heats the upper atmosphere, driving a hydrodynamic outflow. This process is more pronounced for close-in exoplanets around young, active host stars \citep{Ribas}. The observed bimodal radius distribution of sub-Neptunes could have resulted from photoevaporation processes \citep{Fulton2017}.

High-resolution transmission spectroscopy may reveal ongoing atmospheric erosion \citep[e.g.,][]{Spake,Zhang}. During transit, if the planet has an outflowing atmosphere, there will be extra absorption at a few prominent spectral lines such as Lyman-$\alpha$ \citep{Lecavelier}, H$\alpha$ \citep{Feinstein}, and metastable helium \citep{Oklopcic}. This signal can be distinguished from intrinsic variability in the stellar lines because the absorption occurs in the planet’s rest frame. As the planet passes in front of the star, this extra absorption will be blue- and red-shifted by the orbital motion.

\begin{figure*}[htb]
    \centering
    \includegraphics[width=0.7\paperwidth]{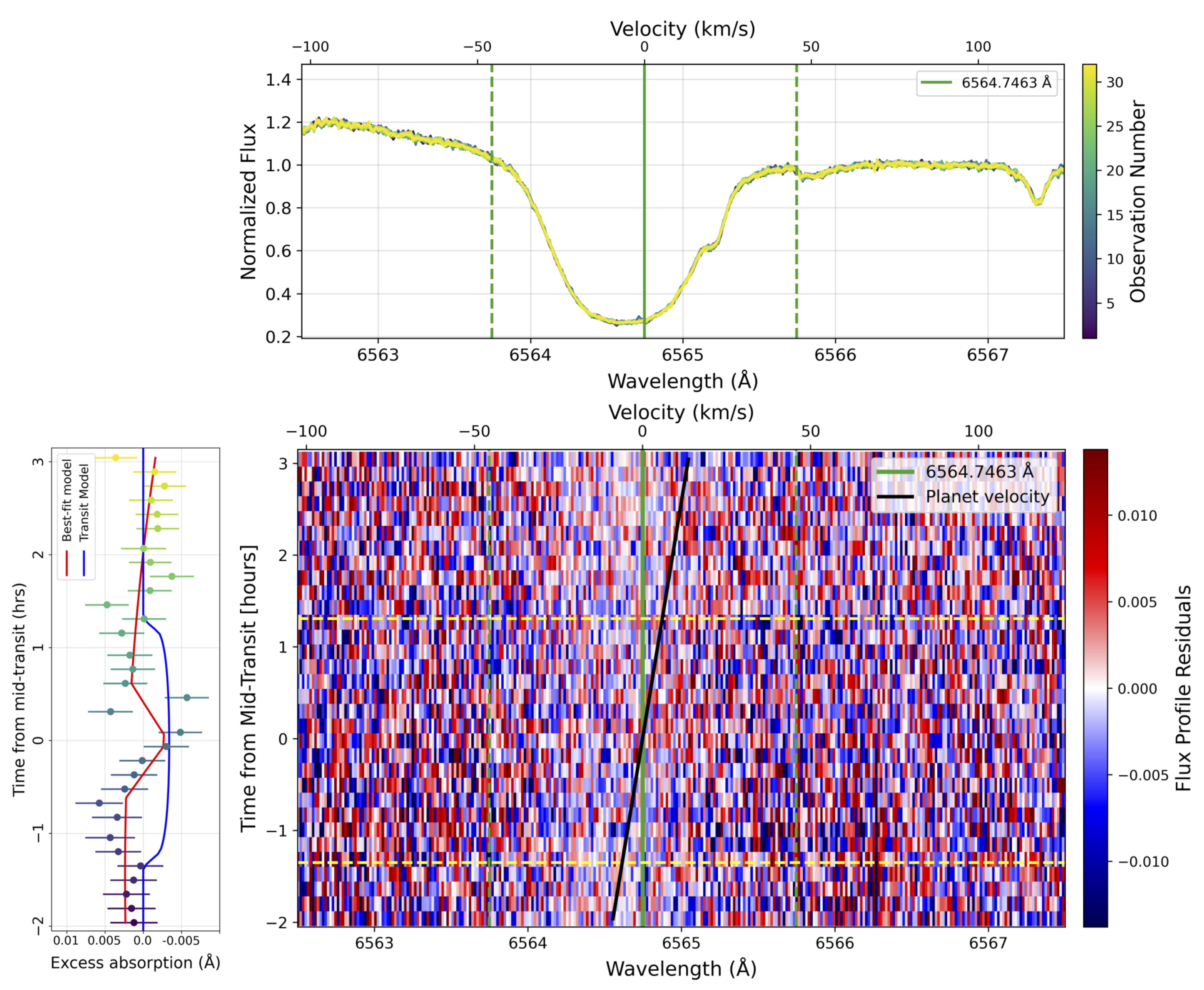}
    \caption{\normalsize Top: KPF H$\alpha$ line during the transit of planet c. There are 32 spectra in total, color-coded by their observation number (epoch). The vertical solid green line marks the line center at 6564.7463\text{\AA}, and the two dashed green lines are 1\text{\AA} away from the center. Left: Excess absorption of H$\alpha$ variation as a function of time, calculated in a bandpass of 2\text{\AA} around the line center. The color scale is the same as the top panel. The red line shows the best-fit trapezoid model with a quadratic baseline (best-fit top width $=0.13 \pm 0.90$ hrs, base width $=1.23 \pm 0.66$ hrs, and height/depth $= 0.004 \pm 0.003$ \text{\AA}), while the blue line represents the white-light-curve transit model of planet c. Bottom right: The measured H$\alpha$ line profile residuals as a function of time and wavelength. The vertical solid and dashed green lines are the same as the top panel. The predicted planetary velocity is marked by the black solid line. The horizontal dashed yellow lines denote the duration of the transit.}
\label{Fig:Mass_loss}
\end{figure*}

Here, we extracted the transmission spectrum of TOI-880 c around the H$\alpha$ line (6564.75\text{\AA}). Figure~\ref{Fig:Mass_loss} shows the KPF line profile during planetary transit and the excess absorption as a function of time and wavelength. To characterize any possible mass loss, we calculated the equivalent width of the H$\alpha$ line as a function of time (the bottom left panel). We constructed a model that is composed of a quadratic baseline function in time and an inverted trapezoidal dip designed to characterize any transient excess absorption feature centered around the transit. We identified the 8th observation as an outlier and excluded it from the fitting process. The resulting best-fit model is shown as the red line. For the trapezoidal model, we obtained a best-fit top width of $0.13 \pm 0.90$ hrs, base width of $1.23 \pm 0.66$ hrs, and height (depth) of $0.004 \pm 0.003$ \text{\AA}. For comparison, the blue line represents the white-light-curve transit model of planet c obtained by {\tt allesfitter}. As evident in the bottom left panel of Figure~\ref{Fig:Mass_loss}, the best-fit trapezoidal model is much shorter in duration than the white-light transit, indicating that the excess absorption does not span the full transit phase. This suggests the signal is unlikely to originate from atmospheric erosion on TOI-880 c. Given the lack of indicators of youth in the host star, photoevaporation on TOI-880 c might have diminished below our detection capability. \citet{Doyle2025} calculated an envelope mass fraction (EMF) of $0.12^{+0.02}_{-0.01}$ for TOI-880 c using the GAS gianT modeL for Interiors \citep[GASTLI;][]{Acuna2021,Acuna2024}. This interior structure model assumes a two-layer configuration consisting of a fully differentiated heavy-element core and a hydrogen--helium envelope, and incorporates the planet’s measured mass, radius, and equilibrium temperature. An EMF of $\sim12\%$ indicates that TOI-880 c has retained a large fraction of its primordial atmosphere. The absence of excess H$\alpha$ absorption in transmission spectroscopy further suggests that the planet is not undergoing vigorous atmospheric escape. However, we defer a numerical upper limit on the mass loss rate to a further study equipped with hydrodynamical models. 

\section{Discussion} \label{sec:discussion}
\subsection{Stellar Obliquity of Multi-Transiting Systems}
\begin{figure*}[htbp]
    \centering
    \includegraphics[width=0.68\textwidth]{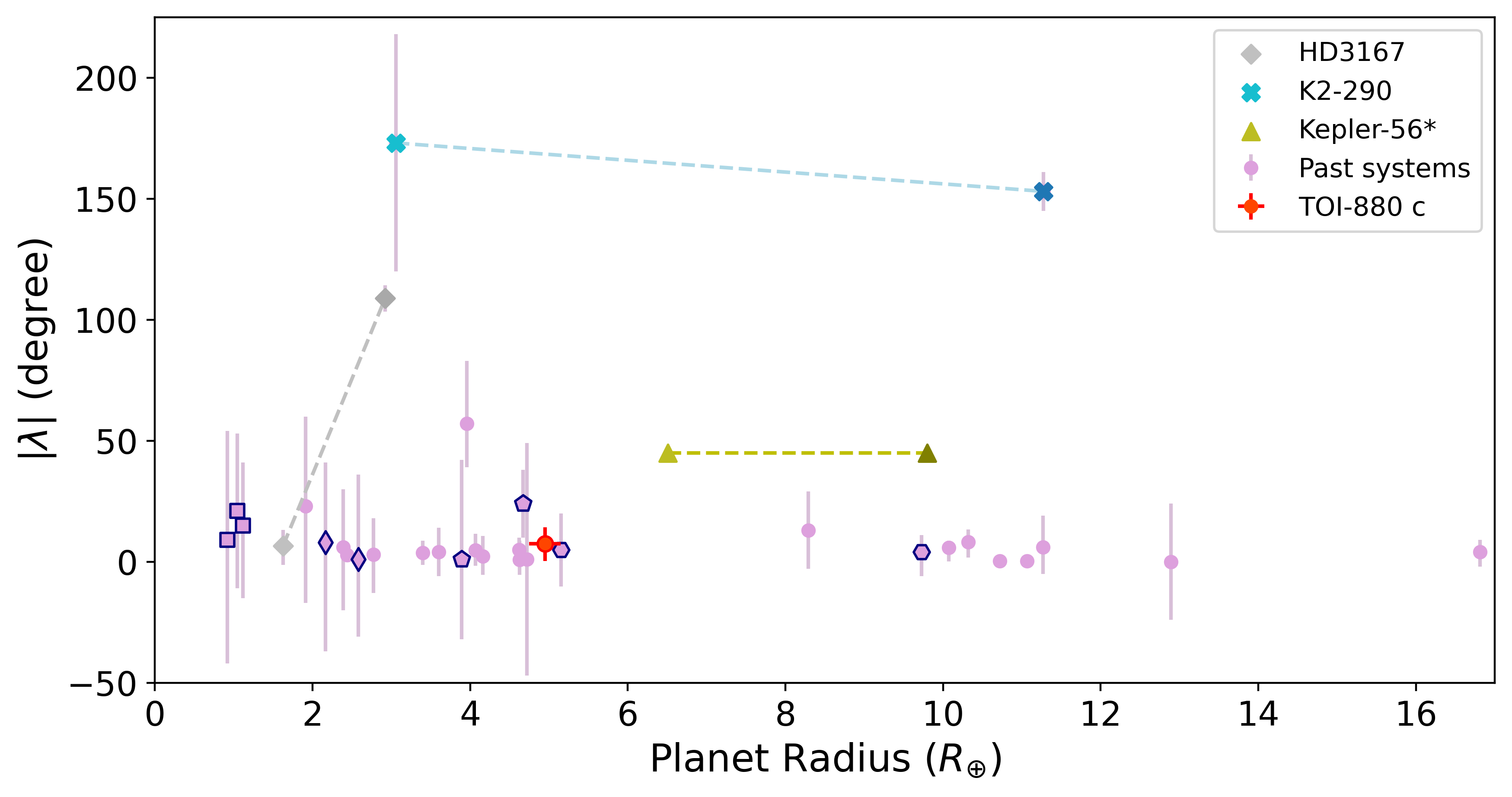}
    \includegraphics[width=0.68\textwidth]{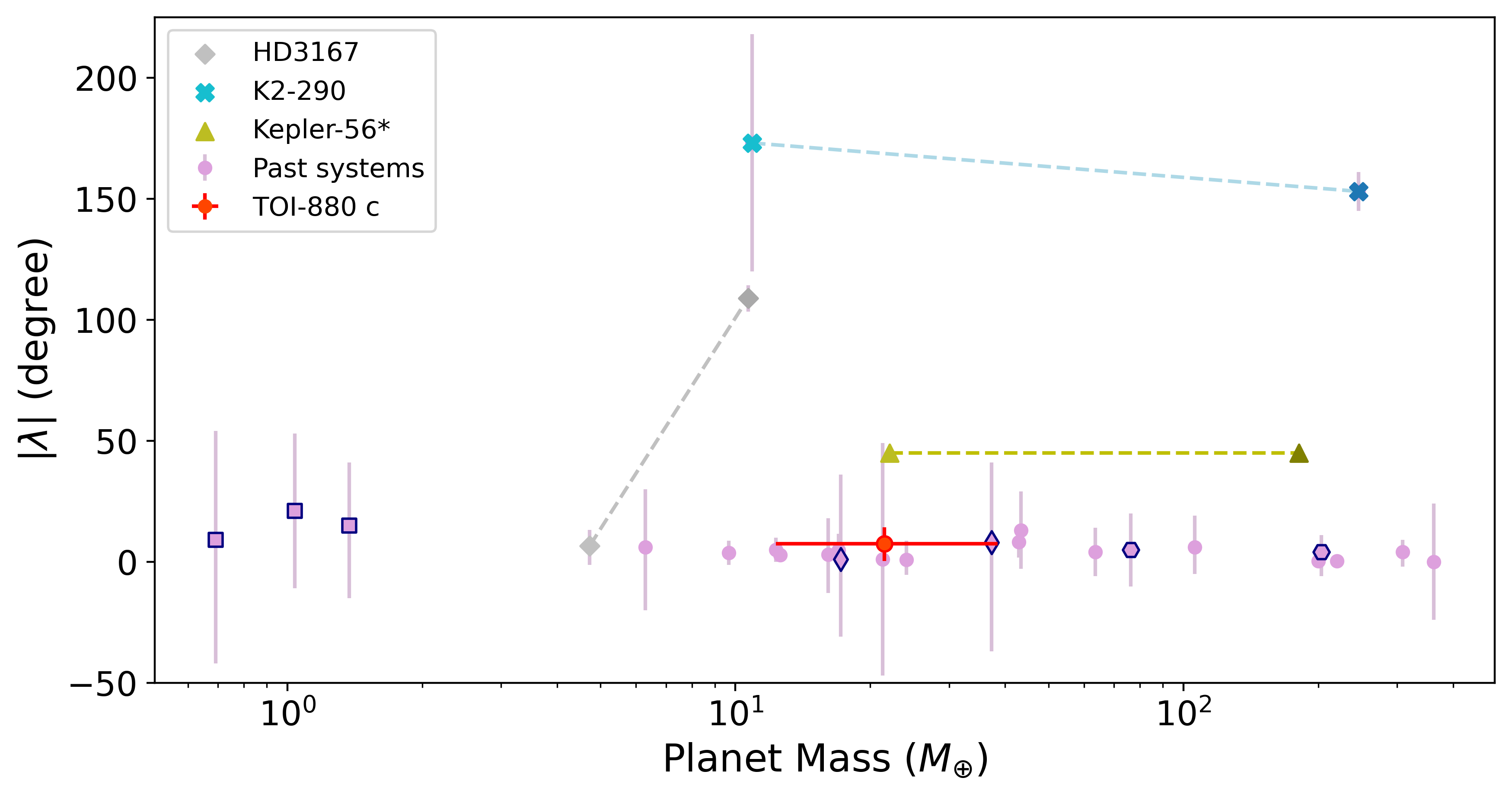}
    \includegraphics[width=0.68\textwidth]{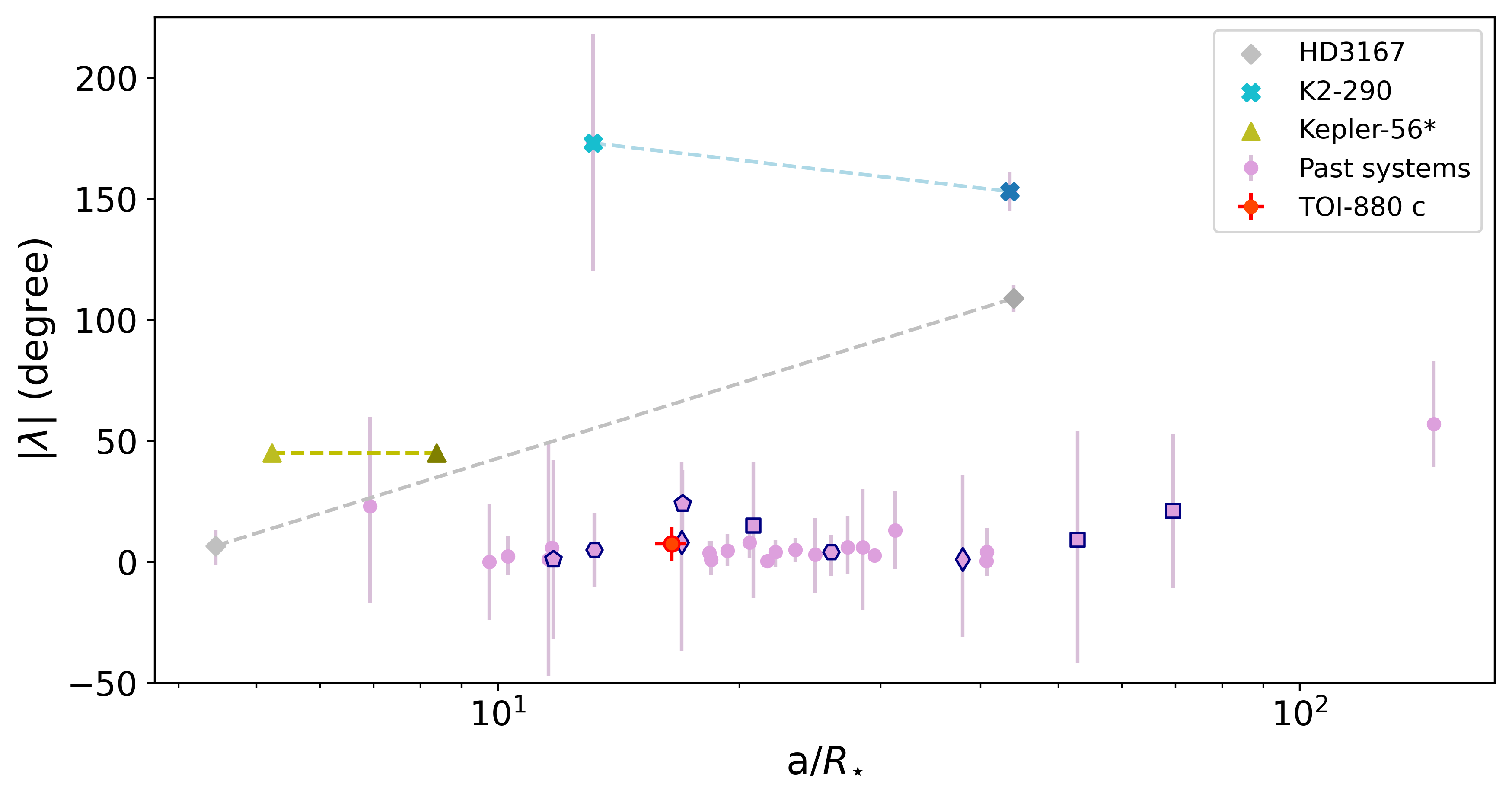}
    \caption{\normalsize The sky-projected obliquity $\lambda$ verses planet radius (top), mass (middle), and semi-major axis over star radius (bottom). Systems that have obliquity measurement using more than one planet are shown as points of the same symbol with dark-blue edges. The three well-known misaligned systems—HD 3167, K2-290, and Kepler-56 \citep[$\ast$which exhibits planet–stellar core misalignment;][]{Kepler-56, Ong2024}—are represented by the connected gray diamonds, blue crosses, and green triangles, respectively. Our target, TOI-880 c, is the red dot with darker edges. It has a fitted sky-projected obliquity of $|\lambda_c| = 7.4_{-7.2}^{+6.8}$$^{\circ}$, radius of $4.95\pm0.20 \mathrm{R_{\oplus}}$, $a/R_{\star}$ of $16.48_{-0.76}^{+0.67}$, and a predicted mass of $21.5^{+17.0}_{-9.2}\,M_{\oplus}$ from \citet{Chen2017}. For the middle panel, only planets with mass measurements are shown.}
\label{Fig:obl_pop}
\end{figure*}

Figure \ref{Fig:obl_pop} shows the sky-projected obliquity $\lambda$ verses planet radius, mass, or semi-major axis over star radius of multi-transiting systems. Other than three well-known misaligned systems, HD 3167 \citep{HD3167}, Kepler-56 \citep[planet-stellar core misalignment;][]{Kepler-56} , and K2-290 \citep{K2-290}, the majority of these systems exhibit low obliquities, consistent with the expectation that multi-transiting planets tend to form and remain in well-aligned configurations. Our system, TOI-880, adds another data point to the growing evidence that compact, multi-planet systems are dynamically cold and largely unaffected by mechanisms that can tilt the whole planetary system \citep{Batygin2012,Lai2011,Rogers2012,Bate,Fielding}. 

 K2-290 A contains two transiting planets both of which have been shown to be on a retrograde orbit around the host star through RM effect or Doppler Shadow \citep{K2-290_2019, K2-290}. Since both planets transit and their RM effects are similar, \citet{K2-290} proposed K2-290 A b and c are coplanar and the common orbital plane is misaligned with the stellar equatorial plane by $124\pm6^{\circ}$. K2-290 A also has two other stellar companions at $\sim$100 (K2-290 B) and $\sim2000$ AU (K2-290 C). \citet{K2-290} suggested that the interplay between the spin and nodal precession (primarily due to K2-290 B) produced a misaligned protoplanetary disk in which K2-290 A b and c formed. \citet{Best} later suggested that the K2-290 C could also be crucial in driving eccentricity and inclination oscillations of the AB orbit, which further chaotic evolution of the stellar obliquities. This could happen after the disk dispersed.

HD 3167 \citep{Vanderburg2016,Gandolfi2017} hosts two transiting planets b and c that was reported to a aligned and polar orbit around the host star with a high mutual inclination $>100^\circ$ \citep{HD3167,Bourrier2021}. However, \citet{Teng2025} found that only about 0.5\% of randomly positioned observers would be able to observe both planets transiting, despite their large mutual inclination (transit strips has minimal overlapping area). Furthermore, in the highly mutually inclined configuration reported for HD 3167 \citep{HD3167,Bourrier2021}, planets b and c would rapidly precess into non-transiting orbits for a fixed observer on century-millennium timescales. We encourage further observation of RM effects on HD-3167 to confirm/rule out the high mutual inclination.

Another misaligned multi-transiting system, Kepler-56 \citep{Kepler-56}, was recently re-analyzed by \citet{Ong2024}. This re-analysis revealed that Kepler-56 may be misaligned with the core rotation of the star but aligned with the surface rotation.

If we leave out HD 3167 in Figure \ref{Fig:obl_pop}, the trend in obliquity of multi-transiting systems is clear. Most compact, multi-transiting systems systems tend to be coplanar and well-aligned, similar to TOI-880. Whereas, K2-290 and Kepler-56 represent a minority population of coplanar but misaligned planetary system. The latter population seems to favor mechanisms that tilted the whole planetary systems such as magnetic warping \citep{Lai2011, Lai2014}, turbulence \citep{Bate2010}, or host star tumbling \citep{Rogers2012, Rogers2013}.

The well-aligned orbit of TOI-880 suggests that the system has likely retained its primordial coplanarity, which is consistent with findings for other compact multi-planet systems \citep{Albrecht2022}. Planetary misalignment can arise from external perturbations or internal disk-planet interactions. External influences, such as distant stellar or planetary companions, can induce obliquities through Kozai-Lidov oscillations \citep{Naoz2011}, leading to periodic inclination and eccentricity changes, which can significantly tilt planetary orbits over Myr timescales. TOI-880 is part of a wide binary system, as identified by \citet{Doyle2025} through cross-matching with Gaia DR3 and the Multiplicity of TESS Objects of Interest catalogue. The stellar companion has a projected separation of approximately 503\,AU and an estimated mass of $0.25\,M_\odot$. TOI-880 has a Gaia RUWE value of 0.996, indicating no evidence for unresolved binarity in the Gaia astrometry. While such wide companions are unlikely to inhibit planet formation, they may influence the orbital architecture of the system through long-term dynamical interactions, such as Kozai--Lidov oscillations or secular perturbations. Despite the presence of a distant stellar companion, the observed spin--orbit alignment in TOI-880 implies that any dynamical interactions—if present—have not significantly altered the system’s coplanarity. This may suggest either that the companion's influence has been dynamically weak due to its wide separation, or that the system experienced a quiescent dynamical history with minimal perturbations. Alternatively, any inclination excitation that did occur may have been damped over time by tidal or disk interactions during the early stages of evolution.

\subsection{JWST Observability}
The James Webb Space Telescope \citep[JWST;][]{Gardner} requires rapid detection, confirmation, and mass measurement of the top atmospheric characterization targets. We found that the transmission spectroscopy metric (TSM) proposed by \citet{Kempton2018} of TOI-880c is $\sim 170$ which makes it a favorable target for JWST observations. Given its coplanarity and alignment, TOI-880 might have remained dynamically quiet since formation.  TOI-880 offers a window into the typical formation and evolution pathway of Kepler-like compact, multi-planet systems.

\section{Summary} \label{sec:summary}
We modeled the KPF RM data produced by TOI-880 c to characterize the stellar obliquity. We derived a sky-projected obliquity of $|\lambda_c| = 7.4_{-7.2}^{+6.8}$$^{\circ}$. Since all three planets transit, the TOI-880 planets are likely coplanar and well-aligned with the host star. The lack of rotational modulation on the host star and low $\vsini$ suggest weak nodal precession due to the host star, so the system will likely remain in this coplanar configuration for extended timescales $\sim$Myr. TOI-880 joins a growing sample of well-aligned, coplanar, multi-transiting systems. Our results also demonstrate the capability of KPF in detecting RM effects for smaller planets, paving the way for future studies on the obliquity distribution of compact multi-planetary systems. We could not robustly detect atmospheric loss on planet c in the H$\alpha$ line, consistent with the host star showing no clear signs of youth.

\bibliography{citation}{}

\begin{thebibliography}{}
\expandafter\ifx\csname natexlab\endcsname\relax\def\natexlab#1{#1}\fi
\providecommand{\url}[1]{\href{#1}{#1}}
\providecommand{\dodoi}[1]{doi:~\href{http://doi.org/#1}{\nolinkurl{#1}}}
\providecommand{\doeprint}[1]{\href{http://ascl.net/#1}{\nolinkurl{http://ascl.net/#1}}}
\providecommand{\doarXiv}[1]{\href{https://arxiv.org/abs/#1}{\nolinkurl{https://arxiv.org/abs/#1}}}

\bibitem[{{Acu{\~n}a} {et~al.}(2021){Acu{\~n}a}, {Deleuil}, {Mousis}, {Marcq}, {Levesque}, \& {Aguichine}}]{Acuna2021}
{Acu{\~n}a}, L., {Deleuil}, M., {Mousis}, O., {et~al.} 2021, \aap, 647, A53, \dodoi{10.1051/0004-6361/202039885}

\bibitem[{{Acu{\~n}a} {et~al.}(2024){Acu{\~n}a}, {Kreidberg}, {Zhai}, \& {Molli{\`e}re}}]{Acuna2024}
{Acu{\~n}a}, L., {Kreidberg}, L., {Zhai}, M., \& {Molli{\`e}re}, P. 2024, \aap, 688, A60, \dodoi{10.1051/0004-6361/202450559}

\bibitem[{{Albrecht} {et~al.}(2022){Albrecht}, {Dawson}, \& {Winn}}]{Albrecht2022}
{Albrecht}, S.~H., {Dawson}, R.~I., \& {Winn}, J.~N. 2022, \pasp, 134, 082001, \dodoi{10.1088/1538-3873/ac6c09}

\bibitem[{{Albrecht} {et~al.}(2021){Albrecht}, {Marcussen}, {Winn}, {Dawson}, \& {Knudstrup}}]{Albrecht2021}
{Albrecht}, S.~H., {Marcussen}, M.~L., {Winn}, J.~N., {Dawson}, R.~I., \& {Knudstrup}, E. 2021, \apjl, 916, L1, \dodoi{10.3847/2041-8213/ac0f03}

\bibitem[{{Barnes} {et~al.}(2013){Barnes}, {van Eyken}, {Jackson}, {Ciardi}, \& {Fortney}}]{Barnes2013}
{Barnes}, J.~W., {van Eyken}, J.~C., {Jackson}, B.~K., {Ciardi}, D.~R., \& {Fortney}, J.~J. 2013, \apj, 774, 53, \dodoi{10.1088/0004-637X/774/1/53}

\bibitem[{{Bate} {et~al.}(2010){Bate}, {Lodato}, \& {Pringle}}]{Bate}
{Bate}, M.~R., {Lodato}, G., \& {Pringle}, J.~E. 2010, \mnras, 401, 1505, \dodoi{10.1111/j.1365-2966.2009.15773.x}

\bibitem[{Bate {et~al.}(2010)Bate, Lodato, \& Pringle}]{Bate2010}
Bate, M.~R., Lodato, G., \& Pringle, J.~E. 2010, Monthly Notices of the Royal Astronomical Society, 401, 1505, \dodoi{10.1111/j.1365-2966.2009.15773.x}

\bibitem[{{Batygin}(2012)}]{Batygin2012}
{Batygin}, K. 2012, \nat, 491, 418, \dodoi{10.1038/nature11560}

\bibitem[{{Beck} \& {Giles}(2005)}]{Beck2005}
{Beck}, J.~G., \& {Giles}, P. 2005, \apjl, 621, L153, \dodoi{10.1086/429224}

\bibitem[{{Best} \& {Petrovich}(2022)}]{Best}
{Best}, S., \& {Petrovich}, C. 2022, \apjl, 925, L5, \dodoi{10.3847/2041-8213/ac49e9}

\bibitem[{{Borucki} {et~al.}(2010){Borucki}, {Koch}, {Basri}, {Batalha}, {Brown}, {Caldwell}, {Caldwell}, {Christensen-Dalsgaard}, {Cochran}, {DeVore}, {Dunham}, {Dupree}, {Gautier}, {Geary}, {Gilliland}, {Gould}, {Howell}, {Jenkins}, {Kondo}, {Latham}, {Marcy}, {Meibom}, {Kjeldsen}, {Lissauer}, {Monet}, {Morrison}, {Sasselov}, {Tarter}, {Boss}, {Brownlee}, {Owen}, {Buzasi}, {Charbonneau}, {Doyle}, {Fortney}, {Ford}, {Holman}, {Seager}, {Steffen}, {Welsh}, {Rowe}, {Anderson}, {Buchhave}, {Ciardi}, {Walkowicz}, {Sherry}, {Horch}, {Isaacson}, {Everett}, {Fischer}, {Torres}, {Johnson}, {Endl}, {MacQueen}, {Bryson}, {Dotson}, {Haas}, {Kolodziejczak}, {Van Cleve}, {Chandrasekaran}, {Twicken}, {Quintana}, {Clarke}, {Allen}, {Li}, {Wu}, {Tenenbaum}, {Verner}, {Bruhweiler}, {Barnes}, \& {Prsa}}]{Borucki}
{Borucki}, W.~J., {Koch}, D., {Basri}, G., {et~al.} 2010, Science, 327, 977, \dodoi{10.1126/science.1185402}

\bibitem[{{Bouma} {et~al.}(2023){Bouma}, {Palumbo}, \& {Hillenbrand}}]{Bouma2023}
{Bouma}, L.~G., {Palumbo}, E.~K., \& {Hillenbrand}, L.~A. 2023, \apjl, 947, L3, \dodoi{10.3847/2041-8213/acc589}

\bibitem[{{Bourrier} {et~al.}(2021){Bourrier}, {Lovis}, {Cretignier}, {Allart}, {Dumusque}, {Delisle}, {Deline}, {Sousa}, {Adibekyan}, {Alibert}, {Barros}, {Borsa}, {Cristiani}, {Demangeon}, {Ehrenreich}, {Figueira}, {Gonz{\'a}lez Hern{\'a}ndez}, {Lendl}, {Lillo-Box}, {Lo Curto}, {Di Marcantonio}, {Martins}, {M{\'e}gevand}, {Mehner}, {Micela}, {Molaro}, {Oshagh}, {Palle}, {Pepe}, {Poretti}, {Rebolo}, {Santos}, {Scandariato}, {Seidel}, {Sozzetti}, {Su{\'a}rez Mascare{\~n}o}, \& {Zapatero Osorio}}]{Bourrier2021}
{Bourrier}, V., {Lovis}, C., {Cretignier}, M., {et~al.} 2021, \aap, 654, A152, \dodoi{10.1051/0004-6361/202141527}

\bibitem[{{Chen} \& {Kipping}(2017)}]{Chen2017}
{Chen}, J., \& {Kipping}, D. 2017, \apj, 834, 17, \dodoi{10.3847/1538-4357/834/1/17}

\bibitem[{{Choi} {et~al.}(2016){Choi}, {Dotter}, {Conroy}, {Cantiello}, {Paxton}, \& {Johnson}}]{MIST}
{Choi}, J., {Dotter}, A., {Conroy}, C., {et~al.} 2016, \apj, 823, 102, \dodoi{10.3847/0004-637X/823/2/102}

\bibitem[{{Coelho} {et~al.}(2005){Coelho}, {Barbuy}, {Mel{\'e}ndez}, {Schiavon}, \& {Castilho}}]{Coelho2005}
{Coelho}, P., {Barbuy}, B., {Mel{\'e}ndez}, J., {Schiavon}, R.~P., \& {Castilho}, B.~V. 2005, \aap, 443, 735, \dodoi{10.1051/0004-6361:20053511}

\bibitem[{{Dalal} {et~al.}(2019){Dalal}, {H{\'e}brard}, {Lecavelier des {\'E}tangs}, {Petit}, {Bourrier}, {Laskar}, {K{\"o}nig}, \& {Correia}}]{HD3167}
{Dalal}, S., {H{\'e}brard}, G., {Lecavelier des {\'E}tangs}, A., {et~al.} 2019, \aap, 631, A28, \dodoi{10.1051/0004-6361/201935944}

\bibitem[{{Dawson} \& {Johnson}(2018)}]{Dawson}
{Dawson}, R.~I., \& {Johnson}, J.~A. 2018, \araa, 56, 175, \dodoi{10.1146/annurev-astro-081817-051853}

\bibitem[{{Doyle} {et~al.}(2025){Doyle}, {Armstrong}, {Acu{\~n}a}, {Osborn}, {Sousa}, {Castro-Gonz{\'a}lez}, {Bourrier}, {Alves}, {Barrado}, {Barros}, {Bayliss}, {Cui}, {Demangeon}, {D{\'\i}az}, {Dumusque}, {Eeles-Nolle}, {Gill}, {Hacker}, {Jenkins}, {Keniger}, {Lafarga}, {Lillo-Box}, {Lockley}, {Nielsen}, {Parc}, {Rodrigues}, {Santerne}, {Santos}, \& {Wheatley}}]{Doyle2025}
{Doyle}, L., {Armstrong}, D.~J., {Acu{\~n}a}, L., {et~al.} 2025, \mnras, 539, 3138, \dodoi{10.1093/mnras/staf670}

\bibitem[{{Eastman} {et~al.}(2019){Eastman}, {Rodriguez}, {Agol}, {Stassun}, {Beatty}, {Vanderburg}, {Gaudi}, {Collins}, \& {Luger}}]{Eastman2019}
{Eastman}, J.~D., {Rodriguez}, J.~E., {Agol}, E., {et~al.} 2019, arXiv e-prints, arXiv:1907.09480, \dodoi{10.48550/arXiv.1907.09480}

\bibitem[{{Fabrycky} {et~al.}(2014){Fabrycky}, {Lissauer}, {Ragozzine}, {Rowe}, {Steffen}, {Agol}, {Barclay}, {Batalha}, {Borucki}, {Ciardi}, {Ford}, {Gautier}, {Geary}, {Holman}, {Jenkins}, {Li}, {Morehead}, {Morris}, {Shporer}, {Smith}, {Still}, \& {Van Cleve}}]{Fabrycky2014}
{Fabrycky}, D.~C., {Lissauer}, J.~J., {Ragozzine}, D., {et~al.} 2014, \apj, 790, 146, \dodoi{10.1088/0004-637X/790/2/146}

\bibitem[{{Feinstein} {et~al.}(2021){Feinstein}, {Montet}, {Johnson}, {Bean}, {David}, {Gully-Santiago}, {Livingston}, \& {Luger}}]{Feinstein}
{Feinstein}, A.~D., {Montet}, B.~T., {Johnson}, M.~C., {et~al.} 2021, \aj, 162, 213, \dodoi{10.3847/1538-3881/ac1f24}

\bibitem[{{Fielding} {et~al.}(2015){Fielding}, {McKee}, {Socrates}, {Cunningham}, \& {Klein}}]{Fielding}
{Fielding}, D.~B., {McKee}, C.~F., {Socrates}, A., {Cunningham}, A.~J., \& {Klein}, R.~I. 2015, \mnras, 450, 3306, \dodoi{10.1093/mnras/stv836}

\bibitem[{{Fulton} {et~al.}(2017){Fulton}, {Petigura}, {Howard}, {Isaacson}, {Marcy}, {Cargile}, {Hebb}, {Weiss}, {Johnson}, {Morton}, {Sinukoff}, {Crossfield}, \& {Hirsch}}]{Fulton2017}
{Fulton}, B.~J., {Petigura}, E.~A., {Howard}, A.~W., {et~al.} 2017, \aj, 154, 109, \dodoi{10.3847/1538-3881/aa80eb}

\bibitem[{{Gaia Collaboration} {et~al.}(2018){Gaia Collaboration}, {Brown}, {Vallenari}, {Prusti}, {de Bruijne}, {Babusiaux}, \& {Bailer-Jones}}]{Gaia}
{Gaia Collaboration}, {Brown}, A.~G.~A., {Vallenari}, A., {et~al.} 2018, ArXiv e-prints.
\newblock \doarXiv{1804.09365}

\bibitem[{{Gaia Collaboration} {et~al.}(2020){Gaia Collaboration}, {Brown}, {Vallenari}, {Prusti}, {de Bruijne}, {Babusiaux}, \& {Biermann}}]{EDR3}
---. 2020, arXiv e-prints, arXiv:2012.01533.
\newblock \doarXiv{2012.01533}

\bibitem[{{Gandolfi} {et~al.}(2017){Gandolfi}, {Barrag{\'a}n}, {Hatzes}, {Fridlund}, {Fossati}, {Donati}, {Johnson}, {Nowak}, {Prieto-Arranz}, {Albrecht}, {Dai}, {Deeg}, {Endl}, {Grziwa}, {Hjorth}, {Korth}, {Nespral}, {Saario}, {Smith}, {Antoniciello}, {Alarcon}, {Bedell}, {Blay}, {Brems}, {Cabrera}, {Csizmadia}, {Cusano}, {Cochran}, {Eigm{\"u}ller}, {Erikson}, {Gonz{\'a}lez Hern{\'a}ndez}, {Guenther}, {Hirano}, {Su{\'a}rez Mascare{\~n}o}, {Narita}, {Palle}, {Parviainen}, {P{\"a}tzold}, {Persson}, {Rauer}, {Saviane}, {Schmidtobreick}, {Van Eylen}, {Winn}, \& {Zakhozhay}}]{Gandolfi2017}
{Gandolfi}, D., {Barrag{\'a}n}, O., {Hatzes}, A.~P., {et~al.} 2017, \aj, 154, 123, \dodoi{10.3847/1538-3881/aa832a}

\bibitem[{{Gardner} {et~al.}(2006){Gardner}, {Mather}, {Clampin}, {Doyon}, {Greenhouse}, {Hammel}, {Hutchings}, {Jakobsen}, {Lilly}, {Long}, {Lunine}, {McCaughrean}, {Mountain}, {Nella}, {Rieke}, {Rieke}, {Rix}, {Smith}, {Sonneborn}, {Stiavelli}, {Stockman}, {Windhorst}, \& {Wright}}]{Gardner}
{Gardner}, J.~P., {Mather}, J.~C., {Clampin}, M., {et~al.} 2006, \ssr, 123, 485, \dodoi{10.1007/s11214-006-8315-7}

\bibitem[{{Gibson} {et~al.}(2016){Gibson}, {Howard}, {Marcy}, {Edelstein}, {Wishnow}, \& {Poppett}}]{Gibson2016}
{Gibson}, S.~R., {Howard}, A.~W., {Marcy}, G.~W., {et~al.} 2016, in Society of Photo-Optical Instrumentation Engineers (SPIE) Conference Series, Vol. 9908, Ground-based and Airborne Instrumentation for Astronomy VI, ed. C.~J. {Evans}, L.~{Simard}, \& H.~{Takami}, 990870, \dodoi{10.1117/12.2233334}

\bibitem[{{Gibson} {et~al.}(2018){Gibson}, {Howard}, {Roy}, {Smith}, {Halverson}, {Edelstein}, {Kassis}, {Wishnow}, {Raffanti}, {Allen}, {Chin}, {Coutts}, {Cowley}, {Curtis}, {Deich}, {Feger}, {Finstad}, {Gurevich}, {Ishikawa}, {James}, {Jhoti}, {Lanclos}, {Lilley}, {Miller}, {Milner}, {Payne}, {Rider}, {Rockosi}, {Sandford}, {Schwab}, {Seifahrt}, {Sirk}, {Smith}, {Stuermer}, {Weisfeiler}, {Wilcox}, {Vandenberg}, \& {Wizinowich}}]{Gibson2018}
{Gibson}, S.~R., {Howard}, A.~W., {Roy}, A., {et~al.} 2018, in Society of Photo-Optical Instrumentation Engineers (SPIE) Conference Series, Vol. 10702, Ground-based and Airborne Instrumentation for Astronomy VII, ed. C.~J. {Evans}, L.~{Simard}, \& H.~{Takami}, 107025X, \dodoi{10.1117/12.2311565}

\bibitem[{{Gibson} {et~al.}(2020){Gibson}, {Howard}, {Rider}, {Roy}, {Edelstein}, {Kassis}, {Grillo}, {Halverson}, {Sirk}, {Smith}, {Allen}, {Baker}, {Beichman}, {Berriman}, {Brown}, {Casey}, {Chin}, {Coutts}, {Cowley}, {Deich}, {Feger}, {Fulton}, {Gers}, {Gurevich}, {Ishikawa}, {James}, {Jelinsky}, {Kaye}, {Lanclos}, {Li}, {Lilley}, {McCarney}, {Miller}, {Milner}, {O'Hanlon}, {Pember}, {Raffanti}, {Rockosi}, {Rubenzahl}, {Rumph}, {Sandford}, {Savage}, {Schwab}, {Seifahrt}, {Shaum}, {Smith}, {Stuermer}, {Thorne}, {Vandenberg}, {Von Boeckmann}, {Wang}, {Wang}, {Weisfeiler}, {Wilcox}, {Wishnow}, {Wizinowich}, {Wold}, \& {Wolfenberger}}]{Gibson2020}
{Gibson}, S.~R., {Howard}, A.~W., {Rider}, K., {et~al.} 2020, in Society of Photo-Optical Instrumentation Engineers (SPIE) Conference Series, Vol. 11447, Ground-based and Airborne Instrumentation for Astronomy VIII, ed. C.~J. {Evans}, J.~J. {Bryant}, \& K.~{Motohara}, 1144742, \dodoi{10.1117/12.2561783}

\bibitem[{{G{\"u}nther} \& {Daylan}(2019)}]{allesfitter-code}
{G{\"u}nther}, M.~N., \& {Daylan}, T. 2019, {Allesfitter: Flexible Star and Exoplanet Inference From Photometry and Radial Velocity}, Astrophysics Source Code Library.
\newblock \doeprint{1903.003}

\bibitem[{{G{\"u}nther} \& {Daylan}(2021)}]{allesfitter-paper}
---. 2021, \apjs, 254, 13, \dodoi{10.3847/1538-4365/abe70e}

\bibitem[{{H{\'e}brard} {et~al.}(2011){H{\'e}brard}, {Ehrenreich}, {Bouchy}, {Delfosse}, {Moutou}, {Arnold}, {Boisse}, {Bonfils}, {D{\'\i}az}, {Eggenberger}, {Forveille}, {Lagrange}, {Lovis}, {Pepe}, {Perrier}, {Queloz}, {Santerne}, {Santos}, {S{\'e}gransan}, {Udry}, \& {Vidal-Madjar}}]{Hebrard2011}
{H{\'e}brard}, G., {Ehrenreich}, D., {Bouchy}, F., {et~al.} 2011, \aap, 527, L11, \dodoi{10.1051/0004-6361/201016331}

\bibitem[{{Higson} {et~al.}(2019){Higson}, {Handley}, {Hobson}, \& {Lasenby}}]{Higson2019}
{Higson}, E., {Handley}, W., {Hobson}, M., \& {Lasenby}, A. 2019, Statistics and Computing, 29, 891, \dodoi{10.1007/s11222-018-9844-0}

\bibitem[{{Hirano} {et~al.}(2011){Hirano}, {Suto}, {Winn}, {Taruya}, {Narita}, {Albrecht}, \& {Sato}}]{Hirano2011}
{Hirano}, T., {Suto}, Y., {Winn}, J.~N., {et~al.} 2011, \apj, 742, 69, \dodoi{10.1088/0004-637X/742/2/69}

\bibitem[{{Hjorth} {et~al.}(2021){Hjorth}, {Albrecht}, {Hirano}, {Winn}, {Dawson}, {Zanazzi}, {Knudstrup}, \& {Sato}}]{K2-290}
{Hjorth}, M., {Albrecht}, S., {Hirano}, T., {et~al.} 2021, Proceedings of the National Academy of Science, 118, e2017418118, \dodoi{10.1073/pnas.2017418118}

\bibitem[{{Hjorth} {et~al.}(2019){Hjorth}, {Justesen}, {Hirano}, {Albrecht}, {Gandolfi}, {Dai}, {Alonso}, {Barrag{\'a}n}, {Esposito}, {Kuzuhara}, {Lam}, {Livingston}, {Montanes-Rodriguez}, {Narita}, {Nowak}, {Prieto-Arranz}, {Redfield}, {Rodler}, {Van Eylen}, {Winn}, {Antoniciello}, {Cabrera}, {Cochran}, {Csizmadia}, {de Leon}, {Deeg}, {Eigm{\"u}ller}, {Endl}, {Erikson}, {Fridlund}, {Grziwa}, {Guenther}, {Hatzes}, {Heeren}, {Hidalgo}, {Korth}, {Luque}, {Nespral}, {Palle}, {P{\"a}tzold}, {Persson}, {Rauer}, {Smith}, \& {Trifonov}}]{K2-290_2019}
{Hjorth}, M., {Justesen}, A.~B., {Hirano}, T., {et~al.} 2019, \mnras, 484, 3522, \dodoi{10.1093/mnras/stz139}

\bibitem[{{Huber} {et~al.}(2013{\natexlab{a}}){Huber}, {Carter}, {Barbieri}, {Miglio}, {Deck}, {Fabrycky}, {Montet}, {Buchhave}, {Chaplin}, {Hekker}, {Montalb{\'a}n}, {Sanchis-Ojeda}, {Basu}, {Bedding}, {Campante}, {Christensen-Dalsgaard}, {Elsworth}, {Stello}, {Arentoft}, {Ford}, {Gilliland}, {Handberg}, {Howard}, {Isaacson}, {Johnson}, {Karoff}, {Kawaler}, {Kjeldsen}, {Latham}, {Lund}, {Lundkvist}, {Marcy}, {Metcalfe}, {Silva Aguirre}, \& {Winn}}]{Kepler-56}
{Huber}, D., {Carter}, J.~A., {Barbieri}, M., {et~al.} 2013{\natexlab{a}}, Science, 342, 331, \dodoi{10.1126/science.1242066}

\bibitem[{{Huber} {et~al.}(2013{\natexlab{b}}){Huber}, {Chaplin}, {Christensen-Dalsgaard}, {Gilliland}, {Kjeldsen}, {Buchhave}, {Fischer}, {Lissauer}, {Rowe}, {Sanchis-Ojeda}, {Basu}, {Handberg}, {Hekker}, {Howard}, {Isaacson}, {Karoff}, {Latham}, {Lund}, {Lundkvist}, {Marcy}, {Miglio}, {Silva Aguirre}, {Stello}, {Arentoft}, {Barclay}, {Bedding}, {Burke}, {Christiansen}, {Elsworth}, {Haas}, {Kawaler}, {Metcalfe}, {Mullally}, \& {Thompson}}]{Huber2013}
{Huber}, D., {Chaplin}, W.~J., {Christensen-Dalsgaard}, J., {et~al.} 2013{\natexlab{b}}, \apj, 767, 127, \dodoi{10.1088/0004-637X/767/2/127}

\bibitem[{{Huber} {et~al.}(2017){Huber}, {Zinn}, {Bojsen-Hansen}, {Pinsonneault}, {Sahlholdt}, {Serenelli}, {Silva Aguirre}, {Stassun}, {Stello}, {Tayar}, {Bastien}, {Bedding}, {Buchhave}, {Chaplin}, {Davies}, {Garc{\'\i}a}, {Latham}, {Mathur}, {Mosser}, \& {Sharma}}]{Huber2017}
{Huber}, D., {Zinn}, J., {Bojsen-Hansen}, M., {et~al.} 2017, \apj, 844, 102, \dodoi{10.3847/1538-4357/aa75ca}

\bibitem[{{Jenkins} {et~al.}(2016){Jenkins}, {Twicken}, {McCauliff}, {Campbell}, {Sanderfer}, {Lung}, {Mansouri-Samani}, {Girouard}, {Tenenbaum}, {Klaus}, {Smith}, {Caldwell}, {Chacon}, {Henze}, {Heiges}, {Latham}, {Morgan}, {Swade}, {Rinehart}, \& {Vanderspek}}]{Jenkins2016}
{Jenkins}, J.~M., {Twicken}, J.~D., {McCauliff}, S., {et~al.} 2016, in Society of Photo-Optical Instrumentation Engineers (SPIE) Conference Series, Vol. 9913, Software and Cyberinfrastructure for Astronomy IV, ed. G.~{Chiozzi} \& J.~C. {Guzman}, 99133E, \dodoi{10.1117/12.2233418}

\bibitem[{{Kempton} {et~al.}(2018){Kempton}, {Bean}, {Louie}, {Deming}, {Koll}, {Mansfield}, {Christiansen}, {L{\'o}pez-Morales}, {Swain}, {Zellem}, {Ballard}, {Barclay}, {Barstow}, {Batalha}, {Beatty}, {Berta-Thompson}, {Birkby}, {Buchhave}, {Charbonneau}, {Cowan}, {Crossfield}, {de Val-Borro}, {Doyon}, {Dragomir}, {Gaidos}, {Heng}, {Hu}, {Kane}, {Kreidberg}, {Mallonn}, {Morley}, {Narita}, {Nascimbeni}, {Pall{\'e}}, {Quintana}, {Rauscher}, {Seager}, {Shkolnik}, {Sing}, {Sozzetti}, {Stassun}, {Valenti}, \& {von Essen}}]{Kempton2018}
{Kempton}, E. M.~R., {Bean}, J.~L., {Louie}, D.~R., {et~al.} 2018, \pasp, 130, 114401, \dodoi{10.1088/1538-3873/aadf6f}

\bibitem[{{Kipping}(2013)}]{Kipping2013}
{Kipping}, D.~M. 2013, \mnras, 435, 2152, \dodoi{10.1093/mnras/stt1435}

\bibitem[{{Lai}(2014)}]{Lai2014}
{Lai}, D. 2014, \mnras, 440, 3532, \dodoi{10.1093/mnras/stu485}

\bibitem[{{Lai} {et~al.}(2011){Lai}, {Foucart}, \& {Lin}}]{Lai2011}
{Lai}, D., {Foucart}, F., \& {Lin}, D. N.~C. 2011, \mnras, 412, 2790, \dodoi{10.1111/j.1365-2966.2010.18127.x}

\bibitem[{{Lecavelier Des Etangs} {et~al.}(2010){Lecavelier Des Etangs}, {Ehrenreich}, {Vidal-Madjar}, {Ballester}, {D{\'e}sert}, {Ferlet}, {H{\'e}brard}, {Sing}, {Tchakoumegni}, \& {Udry}}]{Lecavelier}
{Lecavelier Des Etangs}, A., {Ehrenreich}, D., {Vidal-Madjar}, A., {et~al.} 2010, \aap, 514, A72, \dodoi{10.1051/0004-6361/200913347}

\bibitem[{{Lightkurve Collaboration} {et~al.}(2018){Lightkurve Collaboration}, {Cardoso}, {Hedges}, {Gully-Santiago}, {Saunders}, {Cody}, {Barclay}, {Hall}, {Sagear}, {Turtelboom}, {Zhang}, {Tzanidakis}, {Mighell}, {Coughlin}, {Bell}, {Berta-Thompson}, {Williams}, {Dotson}, \& {Barentsen}}]{lightkurve2018}
{Lightkurve Collaboration}, {Cardoso}, J. V. d.~M., {Hedges}, C., {et~al.} 2018, {Lightkurve: Kepler and TESS time series analysis in Python}, Astrophysics Source Code Library, record ascl:1812.013

\bibitem[{{Lissauer} {et~al.}(2011){Lissauer}, {Ragozzine}, {Fabrycky}, {Steffen}, {Ford}, {Jenkins}, {Shporer}, {Holman}, {Rowe}, {Quintana}, {Batalha}, {Borucki}, {Bryson}, {Caldwell}, {Carter}, {Ciardi}, {Dunham}, {Fortney}, {Gautier}, {Howell}, {Koch}, {Latham}, {Marcy}, {Morehead}, \& {Sasselov}}]{Lissauer2011}
{Lissauer}, J.~J., {Ragozzine}, D., {Fabrycky}, D.~C., {et~al.} 2011, \apjs, 197, 8, \dodoi{10.1088/0067-0049/197/1/8}

\bibitem[{Lomb(1976)}]{Lomb1976}
Lomb, N.~R. 1976, Astrophysics and Space Science, 39, 447, \dodoi{10.1007/BF00648343}

\bibitem[{{McLaughlin}(1924)}]{McLaughlin1924}
{McLaughlin}, D.~B. 1924, \apj, 60, 22, \dodoi{10.1086/142826}

\bibitem[{{McQuillan} {et~al.}(2014){McQuillan}, {Mazeh}, \& {Aigrain}}]{McQuillan2014}
{McQuillan}, A., {Mazeh}, T., \& {Aigrain}, S. 2014, \apjs, 211, 24, \dodoi{10.1088/0067-0049/211/2/24}

\bibitem[{{Naoz} {et~al.}(2011){Naoz}, {Farr}, {Lithwick}, {Rasio}, \& {Teyssandier}}]{Naoz2011}
{Naoz}, S., {Farr}, W.~M., {Lithwick}, Y., {Rasio}, F.~A., \& {Teyssandier}, J. 2011, \nat, 473, 187, \dodoi{10.1038/nature10076}

\bibitem[{{Oklop{\v{c}}i{\'c}} \& {Hirata}(2018)}]{Oklopcic}
{Oklop{\v{c}}i{\'c}}, A., \& {Hirata}, C.~M. 2018, \apjl, 855, L11, \dodoi{10.3847/2041-8213/aaada9}

\bibitem[{{Ong}(2024)}]{Ong2024}
{Ong}, J.~M.~J. 2024, arXiv e-prints, arXiv:2412.19451, \dodoi{10.48550/arXiv.2412.19451}

\bibitem[{{Owen} \& {Wu}(2017)}]{OwenWu2017}
{Owen}, J.~E., \& {Wu}, Y. 2017, \apj, 847, 29, \dodoi{10.3847/1538-4357/aa890a}

\bibitem[{{Petigura}(2015)}]{Petigura_thesis}
{Petigura}, E.~A. 2015, PhD thesis, University of California, Berkeley

\bibitem[{{Petigura} {et~al.}(2017){Petigura}, {Howard}, {Marcy}, {Johnson}, {Isaacson}, {Cargile}, {Hebb}, {Fulton}, {Weiss}, {Morton}, {Winn}, {Rogers}, {Sinukoff}, {Hirsch}, \& {Crossfield}}]{Petigura2017}
{Petigura}, E.~A., {Howard}, A.~W., {Marcy}, G.~W., {et~al.} 2017, \aj, 154, 107, \dodoi{10.3847/1538-3881/aa80de}

\bibitem[{{Pireaux} \& {Rozelot}(2003)}]{Pireaux}
{Pireaux}, S., \& {Rozelot}, J.~P. 2003, \apss, 284, 1159, \dodoi{10.1023/A:1023673227013}

\bibitem[{{Ribas} {et~al.}(2005){Ribas}, {Guinan}, {G{\"u}del}, \& {Audard}}]{Ribas}
{Ribas}, I., {Guinan}, E.~F., {G{\"u}del}, M., \& {Audard}, M. 2005, \apj, 622, 680, \dodoi{10.1086/427977}

\bibitem[{{Rice} {et~al.}(2022){Rice}, {Wang}, \& {Laughlin}}]{Rice2022}
{Rice}, M., {Wang}, S., \& {Laughlin}, G. 2022, \apjl, 926, L17, \dodoi{10.3847/2041-8213/ac502d}

\bibitem[{{Ricker} {et~al.}(2015){Ricker}, {Winn}, {Vanderspek}, {Latham}, {Bakos}, {Bean}, {Berta-Thompson}, {Brown}, {Buchhave}, {Butler}, {Butler}, {Chaplin}, {Charbonneau}, {Christensen-Dalsgaard}, {Clampin}, {Deming}, {Doty}, {De Lee}, {Dressing}, {Dunham}, {Endl}, {Fressin}, {Ge}, {Henning}, {Holman}, {Howard}, {Ida}, {Jenkins}, {Jernigan}, {Johnson}, {Kaltenegger}, {Kawai}, {Kjeldsen}, {Laughlin}, {Levine}, {Lin}, {Lissauer}, {MacQueen}, {Marcy}, {McCullough}, {Morton}, {Narita}, {Paegert}, {Palle}, {Pepe}, {Pepper}, {Quirrenbach}, {Rinehart}, {Sasselov}, {Sato}, {Seager}, {Sozzetti}, {Stassun}, {Sullivan}, {Szentgyorgyi}, {Torres}, {Udry}, \& {Villasenor}}]{Ricker}
{Ricker}, G.~R., {Winn}, J.~N., {Vanderspek}, R., {et~al.} 2015, Journal of Astronomical Telescopes, Instruments, and Systems, 1, 014003, \dodoi{10.1117/1.JATIS.1.1.014003}

\bibitem[{{Rogers} {et~al.}(2012){Rogers}, {Lin}, \& {Lau}}]{Rogers2012}
{Rogers}, T.~M., {Lin}, D.~N.~C., \& {Lau}, H.~H.~B. 2012, \apjl, 758, L6, \dodoi{10.1088/2041-8205/758/1/L6}

\bibitem[{{Rogers} {et~al.}(2013){Rogers}, {Lin}, {McElwaine}, \& {Lau}}]{Rogers2013}
{Rogers}, T.~M., {Lin}, D.~N.~C., {McElwaine}, J.~N., \& {Lau}, H.~H.~B. 2013, \apj, 772, 21, \dodoi{10.1088/0004-637X/772/1/21}

\bibitem[{{Rossiter}(1924)}]{Rossiter1924}
{Rossiter}, R.~A. 1924, \apj, 60, 15, \dodoi{10.1086/142825}

\bibitem[{{Rubenzahl} {et~al.}(2023){Rubenzahl}, {Halverson}, {Walawender}, {Hill}, {Howard}, {Brown}, {Ida}, {Tehero}, {Fulton}, {Gibson}, {Kassis}, {Smith}, {Wold}, \& {Payne}}]{Rubenzahl2023}
{Rubenzahl}, R.~A., {Halverson}, S., {Walawender}, J., {et~al.} 2023, \pasp, 135, 125002, \dodoi{10.1088/1538-3873/ad0b30}

\bibitem[{{Rusznak} {et~al.}(2024){Rusznak}, {Wang}, {Rice}, \& {Wang}}]{Rusznak2024}
{Rusznak}, J., {Wang}, X.-Y., {Rice}, M., \& {Wang}, S. 2024, arXiv e-prints, arXiv:2412.04438, \dodoi{10.48550/arXiv.2412.04438}

\bibitem[{{Scargle}(1982)}]{Scargle1982}
{Scargle}, J.~D. 1982, \apj, 263, 835, \dodoi{10.1086/160554}

\bibitem[{Smith {et~al.}(2012)Smith, Stumpe, Cleve, Jenkins, Barclay, Fanelli, Girouard, Kolodziejczak, McCauliff, Morris, \& Twicken}]{Smith2012}
Smith, J.~C., Stumpe, M.~C., Cleve, J. E.~V., {et~al.} 2012, Publications of the Astronomical Society of the Pacific, 124, 1000, \dodoi{10.1086/667697}

\bibitem[{{Souami} \& {Souchay}(2012)}]{Souami2012}
{Souami}, D., \& {Souchay}, J. 2012, \aap, 543, A133, \dodoi{10.1051/0004-6361/201219011}

\bibitem[{{Spake} {et~al.}(2018){Spake}, {Sing}, {Evans}, {Oklop{\v{c}}i{\'c}}, {Bourrier}, {Kreidberg}, {Rackham}, {Irwin}, {Ehrenreich}, {Wyttenbach}, {Wakeford}, {Zhou}, {Chubb}, {Nikolov}, {Goyal}, {Henry}, {Williamson}, {Blumenthal}, {Anderson}, {Hellier}, {Charbonneau}, {Udry}, \& {Madhusudhan}}]{Spake}
{Spake}, J.~J., {Sing}, D.~K., {Evans}, T.~M., {et~al.} 2018, \nat, 557, 68, \dodoi{10.1038/s41586-018-0067-5}

\bibitem[{{Spalding} \& {Batygin}(2016)}]{Spalding2016}
{Spalding}, C., \& {Batygin}, K. 2016, \apj, 830, 5, \dodoi{10.3847/0004-637X/830/1/5}

\bibitem[{Stumpe {et~al.}(2014)Stumpe, Smith, Catanzarite, Cleve, Jenkins, Twicken, \& Girouard}]{Stumpe2014}
Stumpe, M.~C., Smith, J.~C., Catanzarite, J.~H., {et~al.} 2014, Publications of the Astronomical Society of the Pacific, 126, 100, \dodoi{10.1086/674989}

\bibitem[{{Tayar} {et~al.}(2022){Tayar}, {Claytor}, {Huber}, \& {van Saders}}]{Tayar2022}
{Tayar}, J., {Claytor}, Z.~R., {Huber}, D., \& {van Saders}, J. 2022, \apj, 927, 31, \dodoi{10.3847/1538-4357/ac4bbc}

\bibitem[{{Teng} {et~al.}(2025){Teng}, {Dai}, {Howard}, {Halverson}, {Isaacson}, {Kokubo}, {Rubenzahl}, {Fulton}, {Householder}, {Lubin}, {Giacalone}, {Handley}, {Van Zandt}, {Petigura}, {Ong}, {Premnath}, {Yu}, {Gibson}, {Rider}, {Roy}, {Baker}, {Edelstein}, {Smith}, {Walawender}, {Lee}, {Liu}, \& {Winn}}]{Teng2025}
{Teng}, H.-Y., {Dai}, F., {Howard}, A.~W., {et~al.} 2025, arXiv e-prints, arXiv:2505.10804, \dodoi{10.48550/arXiv.2505.10804}

\bibitem[{{Vanderburg} {et~al.}(2016){Vanderburg}, {Bieryla}, {Duev}, {Jensen-Clem}, {Latham}, {Mayo}, {Baranec}, {Berlind}, {Kulkarni}, {Law}, {Nieberding}, {Riddle}, \& {Salama}}]{Vanderburg2016}
{Vanderburg}, A., {Bieryla}, A., {Duev}, D.~A., {et~al.} 2016, \apjl, 829, L9, \dodoi{10.3847/2041-8205/829/1/L9}

\bibitem[{{Vogt} {et~al.}(2014){Vogt}, {Radovan}, {Kibrick}, {Butler}, {Alcott}, {Allen}, {Arriagada}, {Bolte}, {Burt}, {Cabak}, {Chloros}, {Cowley}, {Deich}, {Dupraw}, {Earthman}, {Epps}, {Faber}, {Fischer}, {Gates}, {Hilyard}, {Holden}, {Johnston}, {Keiser}, {Kanto}, {Katsuki}, {Laiterman}, {Lanclos}, {Laughlin}, {Lewis}, {Lockwood}, {Lynam}, {Marcy}, {McLean}, {Miller}, {Misch}, {Peck}, {Pfister}, {Phillips}, {Rivera}, {Sand ford}, {Saylor}, {Stover}, {Thompson}, {Walp}, {Ward}, {Wareham}, {Wei}, \& {Wright}}]{Vogt2014}
{Vogt}, S.~S., {Radovan}, M., {Kibrick}, R., {et~al.} 2014, \pasp, 126, 359, \dodoi{10.1086/676120}

\bibitem[{{Wang} {et~al.}(2024){Wang}, {Rice}, {Wang}, {Kanodia}, {Dai}, {Logsdon}, {Schweiker}, {Teske}, {Butler}, {Crane}, {Shectman}, {Quinn}, {Kostov}, {Osborn}, {Goeke}, {Eastman}, {Shporer}, {Rapetti}, {Collins}, {Watkins}, {Relles}, {Ricker}, {Seager}, {Winn}, \& {Jenkins}}]{Wang2024}
{Wang}, X.-Y., {Rice}, M., {Wang}, S., {et~al.} 2024, arXiv e-prints, arXiv:2408.10038, \dodoi{10.48550/arXiv.2408.10038}

\bibitem[{{Winn} {et~al.}(2010){Winn}, {Fabrycky}, {Albrecht}, \& {Johnson}}]{Winn2010}
{Winn}, J.~N., {Fabrycky}, D., {Albrecht}, S., \& {Johnson}, J.~A. 2010, \apjl, 718, L145, \dodoi{10.1088/2041-8205/718/2/L145}

\bibitem[{{Wu} \& {Lithwick}(2011)}]{WuLithwick2011}
{Wu}, Y., \& {Lithwick}, Y. 2011, \apj, 735, 109, \dodoi{10.1088/0004-637X/735/2/109}

\bibitem[{{Yu} {et~al.}(2025){Yu}, {Garai}, {Cretignier}, {Szab{\'o}}, {Aigrain}, {Gandolfi}, {Bryant}, {Correia}, {Klein}, {Brandeker}, {Owen}, {G{\"u}nther}, {Winn}, {Heitzmann}, {Cegla}, {Wilson}, {Gill}, {Kriskovics}, {Barrag{\'a}n}, {Boldog}, {Nielsen}, {Billot}, {Lafarga}, {Meech}, {Alibert}, {Alonso}, {B{\'a}rczy}, {Barrado}, {Barros}, {Baumjohann}, {Bayliss}, {Benz}, {Bergomi}, {Borsato}, {Broeg}, {Cameron}, {Csizmadia}, {Cubillos}, {Davies}, {Deleuil}, {Deline}, {Demangeon}, {Demory}, {Derekas}, {Doyle}, {Edwards}, {Egger}, {Ehrenreich}, {Erikson}, {Fortier}, {Fossati}, {Fridlund}, {Gazeas}, {Gillon}, {G{\"u}del}, {Helling}, {Isaak}, {Kiss}, {Korth}, {Lam}, {Laskar}, {Lecavelier des Etangs}, {Lendl}, {Magrin}, {Maxted}, {McCormac}, {Mer{\'\i}n}, {Mordasini}, {Nascimbeni}, {O'Brien}, {Olofsson}, {Ottensamer}, {Pagano}, {Pall{\'e}}, {Peter}, {Piazza}, {Piotto}, {Pollacco}, {Queloz}, {Ragazzoni}, {Rando}, {Rauer}, {Ribas}, {Santos}, {Scandariato}, {S{\'e}gransan}, {Simon}, {Smith}, {Sousa},
  {Southworth}, {Stalport}, {Steinberger}, {Sulis}, {Udry}, {Ulmer}, {Ulmer-Moll}, {Van Grootel}, {Venturini}, {Villaver}, {Walton}, \& {Wheatley}}]{AU_mic_c}
{Yu}, H., {Garai}, Z., {Cretignier}, M., {et~al.} 2025, \mnras, 536, 2046, \dodoi{10.1093/mnras/stae2655}

\bibitem[{{Zhang} {et~al.}(2023){Zhang}, {Knutson}, {Dai}, {Wang}, {Ricker}, {Schwarz}, {Mann}, \& {Collins}}]{Zhang}
{Zhang}, M., {Knutson}, H.~A., {Dai}, F., {et~al.} 2023, \aj, 165, 62, \dodoi{10.3847/1538-3881/aca75b}

\end{thebibliography}
\bibliographystyle{aasjournal}

\end{document}